\documentclass[preprint2]{emulateapj}   

\newcommand\icarus{Icarus}

\shortauthors{Oliveira et al.}
\shorttitle{On the Evolution of Dust Mineralogy}

\begin{document}

\title{On the Evolution of Dust Mineralogy, \\ From Protoplanetary
  Disks to Planetary Systems}

\author{ Isa Oliveira\altaffilmark{1}, Johan Olofsson\altaffilmark{2},
  Klaus M. Pontoppidan\altaffilmark{3,4}, Ewine F. van
  Dishoeck\altaffilmark{1,5}, \\ Jean-Charles Augereau\altaffilmark{6}
  \& Bruno Mer\'{i}n\altaffilmark{7} }
\altaffiltext{1}{Leiden Observatory, Leiden University, P.O. Box 9513,
  2300 RA Leiden, The Netherlands, email: oliveira@strw.leidenuniv.nl}
\altaffiltext{2}{Max-Planck Institut f\"ur Astronomie, K\"onigstuhl 17,
  69117 Heidelberg, Germany}
\altaffiltext{3}{California Institute of Technology, Division for
  Geological and Planetary Sciences, MS 150-21, Pasadena, CA 91125,
  USA}
\altaffiltext{4}{Space Telescope Science Institute, Baltimore, MD
  21218, USA}
\altaffiltext{5}{Max-Planck Institut f\"ur Extraterrestrische Physik,
  Giessenbachstrasse 1, 85748 Garching, Germany}
\altaffiltext{6}{UJF-Grenoble 1/CNRS-INSU, Institut de Plan\'etologie
  et d'Astrophysique de Grenoble (IPAG) UMR 5274, Grenoble, F-38041,
  France}
\altaffiltext{7}{Herschel Science Center, European Space Agency (ESA),
  P.O. Box 78, 28691 Villanueva de la Ca\~nada (Madrid), Spain}

\begin{abstract}

Mineralogical studies of silicate features emitted by dust grains in
protoplanetary disks and Solar System bodies can shed light on the
progress of planet formation. The significant fraction of crystalline
material in comets, chondritic meteorites and interplanetary dust
particles indicates a modification of the almost completely amorphous
interstellar medium (ISM) dust from which they formed. The production
of crystalline silicates, thus, must happen in protoplanetary disks,
where dust evolves to build planets and planetesimals. Different
scenarios have been proposed, but it is still unclear how and when
this happens. This paper presents dust grain mineralogy (composition,
crystallinity and grain size distribution) of a complete sample of
protoplanetary disks in the young Serpens cluster. These results are
compared to those in the young Taurus region and to sources that have
retained their protoplanetary disks in the older Upper Scorpius and
$\eta$ Chamaeleontis stellar clusters, using the same analysis
technique for all samples. This comparison allows an investigation of
the grain mineralogy evolution with time for a total sample of 139
disks. The mean cluster age and disk fraction are used as indicators
of the evolutionary stage of the different populations. Our results
show that the disks in the different regions have similar
distributions of mean grain sizes and crystallinity fractions
($\sim$10 -- 20\%{}) despite the spread in mean ages. Furthermore,
there is no evidence of preferential grain sizes for any given disk
geometry, nor for the mean cluster crystallinity fraction to increase
with mean age in the 1 -- 8 Myr range. The main implication is that a
modest level of crystallinity is established in the disk surface early
on ($\leq$ 1 Myr), reaching an equilibrium that is independent of what
may be happening in the disk midplane. These results are discussed in
the context of planet formation, in comparison with mineralogical
results from small bodies in our own Solar System.

\end{abstract}


\section{Introduction}
\label{sintro}

Protoplanetary disks originate from dense cloud material consisting of
sub-$\mu$m sized, almost completely amorphous interstellar medium
(ISM) dust grains \citep{BE00,LD01,KE04,HG10}. The dust and gas in
these disks form the basic matter from which planets may form. At the
same time, mineralogical studies of primitive solar system bodies
suggest that a considerable fraction of the silicate grains in these
objects are of crystalline nature (\citealt{WO07,PB10}, and references
therein). It is then naturally implied that the crystallinity fraction
increases, through thermal and chemical modification of these solids
during the general planet formation process, commonly referred to in
the literature as ``disk evolution''.

As time passes, the small dust responsible for the infrared (IR)
excess observed around young stars is subjected to different processes
that affect, and will eventually determine how this progression will
end. Planets and planetary systems have been observed around hundreds
of stars other than the Sun, showing that this result is rather common
\citep{US07}. IR observations have revealed a great number of {\it
  debris disks}, composed of large planetesimal rocks and smaller
bodies, around a variety of stars spanning a large range in spectral
types and ages \citep{RI05,BR06,SU06,GA07,CA09}. A few debris disks
are known to harbor planets (e.g., $\beta$ Pictoris and Fomalhaut,
\citealt{LG10,KA08}), although it is still unclear whether this is
often true \citep{KO09}. The majority of main-sequence stars show no
signs of planets or debris within the current observational
limitations, however, indicating that the disks around such stars at
the time of their formation have dissipated completely, leaving no
dust behind to tell the story. Which processes are important and
determinant for the aftermath of disk evolution are still under
debate, and this topic is the subject of many theoretical and
observational studies over the last decade, stimulated in large by
recent IR and (sub-)millimeter facilities.

Specifically on the subject of the mineralogical composition, spectra
from the ground and the {\it Infrared Space Observatory} gave the
first clues of a potential link between crystalline material in
protoplanetary disks and comets. A great similarity was noted between
the spectra of the disk around the Herbig star HD 100546 and that of
comet Hale-Bopp \citep{CR97,MA98}. More recently, the InfraRed
Spectrograph (IRS, 5 -- 38 $\mu$m, \citealt{HO04}) on-board the {\it
  Spitzer Space Telescope} allowed an unprecedented combination of
high sensitivity and the ability to observe large numbers of disks,
down to the brown dwarf limit. The shape of the silicate features
probed by the IRS spectra at 10 and 20 $\mu$m is affected by the
composition, size and structure of its emitting dust. Amorphous
silicates show broad smooth mid-IR features, while the opacities of
crystalline grains show sharp features due to their large-scale
lattice arrangement, such that even small fractions of crystalline
grains produce additional structure in the silicate features
\citep{MI05,BO08,JU09,OF10}. Because most protoplanetary disks are
optically thick at optical and IR wavelengths, the silicate features
observed in the mid-IR are generally emitted by dust in the optically
thin disk surface only. To probe the disk midplane, observations at
longer wavelengths are necessary. Additionally, the emission at 10 and
20 $\mu$m has been shown to arise from different grain populations,
probing different radii \citep{KE06,OF09,OF10}. While the 10 $\mu$m
feature probes a warmer dust population, at $\leq$ 1 AU for T Tauri
stars, the dust emitting at 20 $\mu$m is colder, further out and
deeper into the disk \citep{KE07}.

Two methods have been proposed to explain the formation of crystal
grains: thermal annealing of amorphous grains or vaporization followed
by gas-phase condensation. Both methods require high temperatures
(above $\sim$1000 K, \citealt{FA00,GA04}) which is inconsistent with
outer disk temperatures. However, crystalline grains have been
observed in outer, as well as in inner disks \citep{VB04}. Large-scale
radial mixing has been invoked to explain the presence of crystals at
low temperatures in the outer disk \citep{BM00,GA04,CI09}. A third
proposed formation mechanism for crystal formation is that shock waves
could locally heat amorphous silicates and crystallize them
\citep{DC02,HD02}.

From protoplanetary disks to comets, several authors have attempted to
infer the dust composition from IRS spectra and laboratory data on
amorphous and crystalline silicate dust, using a variety of analysis
techniques. Whether for individual objects
\citep{FO04,ME07,PI08,BY08}, for mixed disk samples
\citep{BO01,AP05,VB05,BO08,OF09,OF10,JU10}, or systematic studies of
the disk population of a given star-forming region
\citep{SI09,WA09,ST09}, it has been shown that a significant mass
fraction of the dust in those disks must be in crystalline
form. However, the many studies dealing with the mineralogical
composition of dust to date focus on a specific region or object,
failing to investigate the hypothesis that the crystallinity fraction
is a measure of the evolutionary stage of a region. That is, no study
in the literature has yet investigated an increase of crystallinity
fraction with cluster age.

Mineralogical studies of Solar System bodies show a range of
crystallinity fractions. Evidence from primitive chondrites shows that
the abundance of crystalline silicate material varies from nearly
nothing up to 20 -- 30~\% (e.g. Acfer 094 and ALH77307, \citealt{PB10}
and references therein). Oort cloud comets, with long periods and
large distances from the Sun, have inferred crystallinity fractions up
to 60 -- 80 \% (e.g. Hale-Bopp, \citealt{WO99,WO07}). Jupiter-family,
or short period comets, have lower fractions, up to $\sim$35 \%
(e.g. 9P/Tempel 1, \citealt{HA07}; 81P/Wild 2, \citealt{ZO06}). This
discrepancy in fractions points to the existence of a radial
dependence in crystallinity fraction in the protoplanetary disk around
the young Sun \citep{HA05}. It is important to note that those values
are model dependent, and the use of large amorphous grains (10 -- 100
$\mu$m) can lead to systematically lower crystalline fractions
\citep{HA02}. This is evident for Hale-Bopp, where \citet{MI05a} find
a much lower fraction ($\sim$7.5 \%) than other authors, using a
distribution of amorphous grain sizes up to 100 $\mu$m. What is clear
is that even within the discrepancies, the crystallinity fractions
derived for Solar System bodies are appreciably higher than those
derived for the ISM dust ($< 2 \%$, \citealt{KE04}). Recent {\it
  Spitzer} data indicate further similarities between crystalline
silicate features seen in comets or asteroids with those seen in some
debris disks around solar mass stars \citep{BE06,LI07,LI08}. One
proposed explanation is that the observed spectral features in the
disk result from the catastrophic break-up of a single large body (a
`super comet') which creates the small dust particles needed for
detection. At the even earlier protoplanetary disk stage, there is
limited observational evidence for radial gradients in crystallinity
from mid-infrared interferometry data, with higher crystallinity
fractions found closer to the young stars \citep{VB04,SC08}. All of
this suggests that the crystallization occurs early in the disk
evolution and is then incorporated into larger solid bodies.

Besides dust composition, the evolution of grain sizes is an essential
indicator of disk evolution. The initially sub-$\mu$m size ISM grains
must grow astounding 14--15 orders of magnitude in diameter if they
are to form planets. If grains were to grow orderly and steadily,
theoretical calculations predict disks to have fully dissipated their
small grains within $\sim$10$^5$ years \citep{WE80,DD05}. The fact
that many disks a few Myr old are observed to have small grains
\citep{HE08} poses a serious problem for the paradigm that grain
growth is a steady, monotonic process in disk evolution and planet
formation. Additionally, small dust has been observed in the surface
layers of disks in clusters of different ages and environments for
hundreds of systems. The implications, as discussed most recently by
\citet{OL10} and \citet{OF10}, is that small grains must be
replenished by fragmentation of bigger grains, and that an equilibrium
between grain growth and fragmentation is established. \citet{OL10}
have shown that this equilibrium is maintained over a few million
years, as long as the disks are optically thick, and is independent of
the population or environment studied.

In this paper we present a comprehensive study of the mineralogical
composition of disks around stars in young star-forming regions (where
most stars are still surrounded by optically thick disks) and older
clusters (where the majority of disks has already dissipated).
Correlating the results on mean size and composition of dust grains
per region, obtained in a homogeneous way using the same methodology,
with the properties of small bodies in our own Solar System can put
constraints on some of the processes responsible for disk evolution
and planet formation. The Serpens Molecular Cloud, whose complete
flux-limited YSO population has been observed by the IRS instrument
\citep{OL10}, is used as a prototype of a young star-forming region,
together with Taurus, the best studied region to date. The sources
that have retained their protoplanetary disks in the $\eta$
Chamaeleontis and Upper Scorpius clusters are used to probe the
mineralogy in the older bin of disk evolution.

Section~\ref{sdata} describes the YSO samples in the 4 regions
mentioned. The {\it Spitzer} IRS observations and reduction are
explained. The spectral decomposition method B2C \citep{OF10} is
briefly introduced in \S~\ref{s_spdecomp}, and its results for
individual and mean cluster grain sizes and composition are shown in
\S~\ref{sres}. In \S~\ref{sdis} the results are discussed in the
context of time evolution. There we demonstrate that no evolution is
seen in either mean grain sizes or crystallinity fractions as clusters
evolve from $\sim$1 to 8 Myr. The implications for disk formation and
dissipation, and planet formation are discussed. In \S~\ref{scon} we
present our conclusions.

\section{Spitzer IRS data}
\label{sdata}

The four regions presented here were chosen due to the availability of
complete sets of IRS spectra of their IR-excess sources, while
spanning a wide range of stellar characteristics, environment, mean
ages and disk fractions (the disk fraction of Serpens is still
unknown, see Table \ref{t_overview}).

The IRS spectra of a complete flux-limited sample of young stellar
objects (YSO) in the Serpens Molecular Cloud have been presented by
\citet{OL10}, based on program ID \#30223 (PI: Pontoppidan). As
detailed there, the spectra were extracted from the basic calibration
data (BCD) using the reduction pipeline from the Spitzer Legacy
Program ``From Molecular Cores to Planet-Forming Disks'' (c2d,
\citealt{LH06}). A similarly large YSO sample in the Taurus
star-forming region has been presented by \citet{FU06}. IRS spectra of
all 18 members of the $\eta$ Chamaeleontis cluster were first shown by
\citet{SI09}, while the spectra of 26 out of the 35 IR-excess sources
in the Upper Scorpius OB association were shown by \citet{DC09} (the
remaining 9 objects were not known at the time the observations were
proposed). For the latter 3 regions, the post-BCD data were downloaded
from the SSC pipeline (version S18.4) and then extracted with the
Spitzer IRS Custom Extraction software (SPICE, version 2.3) using the
batch generic template for point sources. As a test, the IRS spectra
of the YSOs in Serpens were also reduced using SPICE to ensure that
both pipelines produce nearly identical results. On visual inspection,
no discrepancies were found between the results from the two
pipelines, all objects showed the exact same features in both
spectra. The similarity in outputs is such that the effects on the
spectral decomposition results are within the cited error bars.

Since the spectral decomposition method applied here aims to reproduce
the silicate emission from dust particles in circumstellar disks, the
sample has been limited to spectra that show clear silicate features.
The few sources with PAH emission have been excluded from the
sample. PAH sources amount to less than 8\% in low-mass star-forming
regions \citep{GE06,OL10}. Furthermore, spectra with very low
signal-to-noise ratios (S/N) are excluded from the analysed sample in
order to guarantee the quality of the results. In addition, for
objects \#114 and 137 in Serpens, and 04370+2559 and V955Tau in Taurus
the warm component fit contributes to most of the spectrum, leaving
very low fluxes to be fitted by the cold component. This produces
large uncertainties in the cold component fit, and they are therefore
not further used in the analysis. The low S/N objects rejected amount
to less than 10~\% of each of the Serpens and Taurus samples, so the
statistical results derived here should not be affected by this
removal. The final sample of 139 sources analysed is composed of 60
objects in Serpens, 66 in Taurus, 9 objects in Upper Scorpius, and 4
in $\eta$ Chamaeleontis. The statistical uncertainties of the spectra
were estimated as explained in \citet{OF09}.

The great majority of the objects studied here are low-mass stars
(spectral types K and M, see Table \ref{t_comp}). The study of
mineralogical evolution across stellar mass is not the focus of this
paper. Such a study would require a separate paper, in which the same
techniques are used for low- and intermediate-mass stars. Thus, the
statistical results derived in the following sections concern T Tauri
stars, and not necessarily apply to intermediate-mass Herbig Ae/Be
stars.

\section{Spectral Decomposition and the B2C method}
\label{s_spdecomp}

\begin{figure}[!h]
\begin{center}
\includegraphics[width=0.4\textwidth]{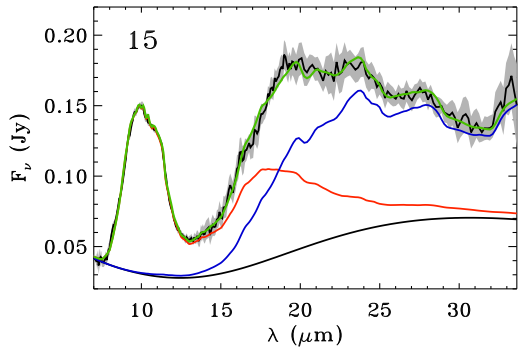}
\end{center}
\caption{\label{f_fit} Example of the B2C modeling for object \#15 in
  Serpens. The black line is the estimated continuum for this
  source. The red line is the fit to the warm component and the blue
  line is the fit to the cold component. The green line is the final
  fit to the entire spectrum. The original spectrum is shown in black
  with its uncertainties in light grey.}
\end{figure}

In order to reproduce the observed IRS spectra of these circumstellar
disks the B2C decomposition method, explained in detail and tested
extensively in \citet{OF10}, is applied. Two dust grain populations,
or components, at different temperatures (warm and cold) are used in
the method, in addition to a continuum emission. The warm component
reproduces the 10 $\mu$m feature, while the cold component reproduces
the non-negligible residuals at longer wavelengths, over the full
spectral range (see Figure \ref{f_fit}). Each component, warm and
cold, is the combination of five different dust species and three
grain sizes for amorphous silicates or two grain sizes for crystalline
silicates.

The three amorphous species are silicates of olivine stoichiometry
(MgFeSiO$_4$), silicates of pyroxene stoichiometry (MgFeSiO$_6$), and
silica (SiO$_2$). The two crystalline species are both Mg-rich end
members of the pyroxene and olivine groups, enstatite (MgSiO$_3$) and
forsterite (Mg$_2$SiO$_4$). As further explained in \citet{OF10}, the
theoretical opacities of the amorphous species are computed assuming
homogeneous spheres (Mie theory), while those for the crystalline
species use the distribution of hollow spheres (DHS, \citealt{MI05})
theory so that irregularly shaped particles can be simulated.

In addition, the three grain sizes used are 0.1, 1.5 and 6.0 $\mu$m,
representing well the spectroscopic behaviour of very small,
intermediate-sized and large grains. For the crystalline species,
however, the code is limited to only 2 grain sizes (0.1 and 1.5
$\mu$m). This restriction is imposed because large crystalline grains
are highly degenerate with large amorphous grains (as can be seen in
Figure 1 of \citealt{OF10}), and because the production of large 6.0
$\mu$m pure crystals is not expected via thermal annealing
\citep{GA04}.

The B2C method itself consists of three steps. First, the continuum is
estimated and subtracted from the observed spectrum. The adopted
continuum is built by using a power-law plus a black-body at
temperature $T_{\rm cont}$. The power-law represents the mid-IR tail
of emission from the star and inner disk rim. The black-body is
designed to contribute at longer wavelengths, and is therefore
constrained to be less than 150 K. Each dust component is then fitted
separately to the continuum-subtracted spectrum.

The second step is to fit the warm component to reproduce the 10
$\mu$m silicate feature between $\sim$7.5 and 13.5 $\mu$m. This is
done by summing up the 13 mass absorption coefficients ($N_{\rm
  species}$ = 5, $N_{\rm sizes}$ = 3 or 2, for amorphous and
crystalline species, respectively), multiplied by a black-body
$B_{\nu} (T_{\rm w})$ at a given warm temperature $T_{\rm w}$.

The third step is to fit the residuals, mostly at longer wavelengths,
over the entire spectral range (5 -- 35 $\mu$m). This is done in a
similar manner, for a given cold temperature $T_{\rm c}$. The final
fit is a sum of the three fits described, as can be seen in Figure
\ref{f_fit}. The entire fitting process is based on a Bayesian
analysis, combined with a Monte Carlo Markov chain, in order to
randomly explore the space of free parameters. The resulting mean
mass-average grain size is the sum of all sizes fitted, each size
being weighted by their corresponding masses, as:

\begin{tiny}
\begin{eqnarray}
  \langle a_{\rm warm/cold} \rangle = 
  \left(\sum_{j=1}^{\mathrm{N_{\rm sizes}}}  a_{j} 
    \sum_{i=1}^{\mathrm{N_{\rm species}}}  
    M_{{\rm w/c},i}^{j} \right)
  \times \left( \sum_{j=1}^{\mathrm{N_{\rm sizes}}} 
    \sum_{i=1}^{\mathrm{N_{\rm species}}}  
    M_{{\rm w/c},i}^{j} \right)^{-1}
  \label{eq:meana}
\end{eqnarray}
\end{tiny}
where $a_1 = 0.1\,\mu$m (small grains), $a_2 = 1.5\,\mu$m
(intermediate-sized grains) and $a_3 = 6\,\mu$m (large grains).
Further details and tests of the B2C procedure can be found in
\citet{OF10}. That paper also demonstrates that the procedure is
robust for statistical samples, and that the relative comparisons
between samples, which are the focus of this paper, should not suffer
from the assumptions that enter in the procedure. The robustness of
the procedure is evaluated by fitting synthetic spectra, and is
discussed in detail in their Appendix A. The influence of the
continuum estimate is also discussed, especially for the cold
component for both grain sizes and crystallinity fractions, and it is
shown that prescriptions that do not use large 6 $\mu$m grains (which
are, to some degree, degenerate with the continuum) give fits that are
not so good.

For the amorphous grains, the B2C procedure uses the Mie scattering
theory to compute mass absorption coefficients. However, \citet{MI07}
found that they could best reproduce the extinction profile toward the
galactic center using the DHS scattering theory, with a maximum
filling factor of 0.7. The most striking difference between Mie and
DHS mass absorption coefficients is seen for the O--Si--O bending mode
around 20 $\mu$m. Here we investigate the influence of the use of DHS
instead of Mie for amorphous grain with an olivine or pyroxene
stoichiometry. We conducted tests on a sub-sample of 30 objects (15 in
Serpens and 15 in Taurus). The conclusion of such tests is that it has
a small influence on the quantities we discuss in this study. For the
warm component of the 30 objects, we find a change in the mean
crystallinity fraction of -1.6\% (the mean crystallinity for this
sub-sample using DHS is 9.3\% versus 10.9\% using Mie), which is in
the range of uncertainties claimed in this study. We also computed the
mean slope of grain size distributions to gauge the effect of using
DHS on grain sizes. On average, the grain size distribution indices
are steeper by $\sim$0.2 (with a mean slope of -3.01 for this
sub-sample using DHS versus -2.80 using Mie). Therefore, our main
conclusions are preserved for the warm component. Concerning the cold
component, the inferred crystallinity fraction using DHS is 22.5\%
versus 15.1\% with Mie, a mean increase of 7.4\%. For the mean slope
of grain size distributions, a negligible decrease is found (-3.07
using DHS versus -3.01 for Mie). Again, the differences found are
within our significant errors for the cold component and do not change
any of our conclusions.

It is important to note that the S/N generally degrades at longer
wavelengths when compared to shorter wavelengths. The lower S/N
reflect on the cold component fits and will most likely result in
larger uncertainties. We evaluate that the fits to the cold component
are reliable and add important information on the dust mineralogy
(albeit with larger uncertainties) and thus those results are included
in the following discussion.

\section{Results}
\label{sres}

The IRS spectra of the 139 YSOs with IR excess discussed in
\S~\ref{sdata} were fitted with the B2C spectral decomposition
procedure. The relative abundances derived for all objects are shown
in Appendix \ref{sabun}. The S/N drops considerably for the long
wavelength module of some of the objects studied (including all
objects in Upper Scorpius and $\eta$ Chamaeleontis). For this reason,
the cold component could not be satisfactorily fitted and no results
for this component are presented for these sources (see Appendix
\ref{sabun}).

Due to the large number of objects, these results allow statistical
studies on both the mineralogy and size distribution of the grains
that compose the optically thin surface layers of disks in each
cluster studied. The mean abundances of each species per region are
presented in Table \ref{t_mean_comp}, where it can be seen that the
majority of the dust studied is of amorphous form. In Table
\ref{t_mean} the mean mass-average grain sizes and crystallinity
fractions per region are shown. Mean sizes are in the range 1 -- 3
$\mu$m, without significant difference between regions. These results
are discussed in detail in the following sections.

It is important to note that the comparison of results derived here
for the different regions is valid because the same method, with exact
same species, is used for all sources. The comparison of samples
analyzed in distinct ways can lead to differences in results that do
not correspond to real differences in composition. Nevertheless, in
\S~\ref{scomp} the results presented here are compared to literature
results for the same objects, when available, with generally good
agreement.

\subsection{Grain Sizes}
\label{s_gsize}

\begin{figure}[!h]
\begin{center}
\includegraphics[width=0.35\textwidth]{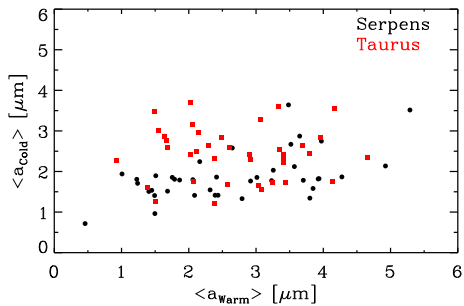}
\end{center}
\caption{\label{f_size} Mass-averaged mean grain sizes for the warm
  ($\langle a_{\rm warm} \rangle$) and cold ($\langle a_{\rm cold}
  \rangle$) components. Black dots are the objects in Serpens, and red
  squares are the objects in Taurus.}
\end{figure}

The mean mass-averaged grain sizes for the warm ($\langle a_{\rm warm}
\rangle$) and cold ($\langle a_{\rm cold} \rangle$) components are
shown in Figure \ref{f_size}, for Serpens and Taurus (for the objects
in Upper Sco and $\eta$ Cha no results for the cold component are not
available, see Appendix \ref{sabun}). It is seen that the two clouds
overlap greatly, and that the grain sizes derived from the different
temperature components do not seem to correlate. To quantify this
correlation, a Kendall $\tau$ correlation coefficient can be computed
together with its associated probability $P$ (between 0 and 1). $\tau$
= 1(-1) defines a perfect correlation (anti-correlation), and $\tau$ =
0 means that the datasets are completely independent. A small $P$, on
the other hand, testifies to how tight the correlation is. For the
warm and cold mean mass-averaged grain sizes for both clouds, $\tau$
is found to be 0.14, with $P$ = 0.07. This lack of correlation
indicates that different processes are likely responsible for
regulating the size distribution at different radii \citep{OF10}.

\begin{figure}[!h]
\begin{center}
\includegraphics[width=0.45\textwidth]{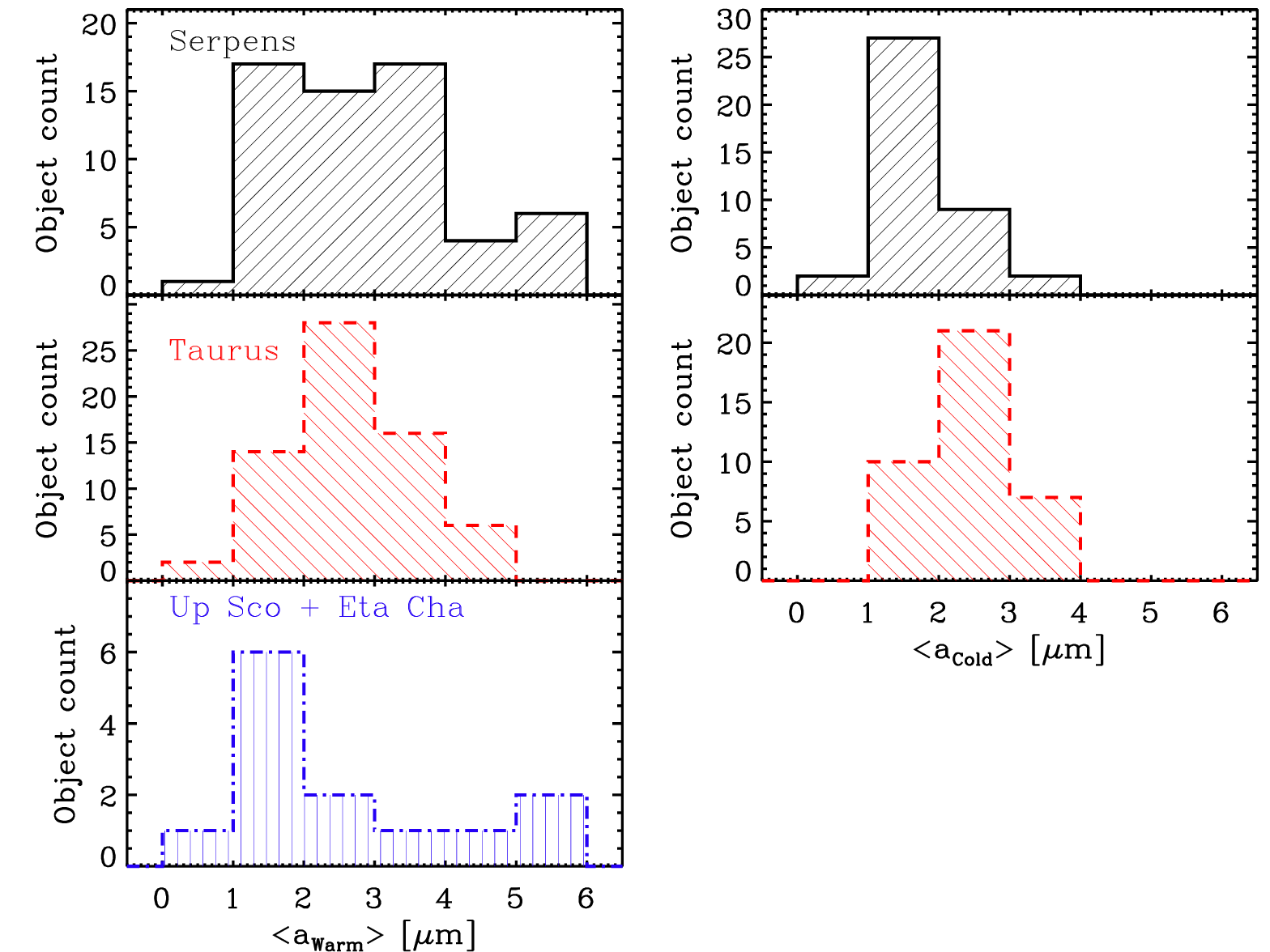}
\end{center}
\caption{\label{f_sizedist} Distribution of mass-averaged mean grain
  sizes for the warm ($\langle a_{\rm warm} \rangle$, left panel) and
  cold ($\langle a_{\rm cold} \rangle$, right panel) components. Due
  to the low number statistics, the objects in Upper Sco and $\eta$
  Cha have been merged together as an older cluster.}
\end{figure}

Although the average grain size in the warm component is bigger than
that in the cold component within a given star-forming region, as
shown in Table \ref{t_mean}, this difference is mostly not
significant. However, Figures \ref{f_size} and \ref{f_sizedist}
clearly show a difference between the range of grain sizes spanned in
both components, with $\langle a_{\rm cold} \rangle$ never reaching
near the biggest grain size modeled (6.0 $\mu$m) for any object. A
possible explanation for larger grains at smaller radii, suggested by
\citet{ST09}, is that grains coagulate faster in the inner disk where
dynamical timescales are shorter.  However as discussed by
\citet{OL10} and in \S~\ref{sover}, the mean dust size at the disk
surface is not regulated by grain growth alone, but also by
fragmentation and vertical mix. This means that faster coagulation at
smaller radii cannot be uniquely responsible for bigger grains in the
inner disk. Future modeling should try to understand this difference
in mean grain sizes observed.

Furthermore, Serpens and Taurus occupy an indistinguishable locus in
Figure \ref{f_size}, explicitly seen in Figure \ref{f_sizedist}. A two
sample Kolmogorov-Smirnov test (KS-test) was performed and the results
show that the null hypothesis that the two distributions come from the
same parent population cannot be rejected to any significance
(14\%). The older regions, although lacking statistical significance,
show a distribution of mass-average grain sizes in the same range
probed by the young star-forming regions (Figure
\ref{f_sizedist}). This supports the evidence that the size
distribution of the dust in the surface layers of disks is
statistically the same independent of the population studied
\citep{OL10}. 

The results here confirm those from \citet{OF10} that the mean
differential grain size distributions slope for the three grain sizes
considered are shallower than the reference MRN differential size
distribution ($\alpha$ = -3.5). The mean grain size distributions
slopes ($\alpha$) for each region can be found in Table \ref{t_mean}.

\subsection{Disk Geometry}
\label{s_dgeo}

The amount of IR-excess in a disk is directly related to its geometry
\citep{KE87,GM01,DU01}. Specifically using the IRS spectra, disk
geometry can be inferred from the flux ratio between 30 and 13 $\mu$m
($F_{30}/F_{13}$, \citealt{JB07,OL10,ME10}). A flared geometry ($1.5
\lesssim F_{30}/F_{13} \lesssim 5$), with considerable IR excess and
small dust, allows the uppermost dust layers to intercept stellar
light at both the inner and outer disk. For flat disks ($F_{30}/F_{13}
\lesssim 1.5$) with little IR excess, only the inner disk can easily
intercept the stellar radiation as the outer disk is
shadowed. Moreover, cold or transitional disks are interesting objects
that present inner dust gaps or holes, producing a region with little
or no near-IR excess ($5 \lesssim F_{30}/F_{13} \lesssim 15$). It is
interesting to explore the effect of disk geometry on both the mean
mass-average grain sizes and crystallinity fractions of the disks
studied.

\begin{figure}[h]
\begin{center}
\includegraphics[width=0.45\textwidth]{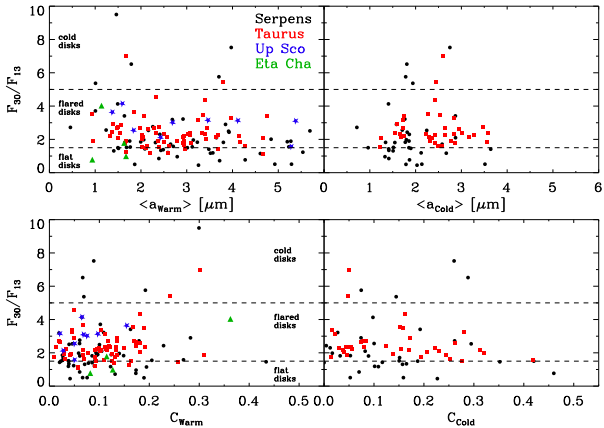}
\end{center}
\caption{\label{f_f30} Top: Flaring index $F_{30}/F_{13}$, used as a
  proxy for disk geometry, versus warm (left) and cold (right)
  mass-averaged mean grain sizes. Bottom: $F_{30}/F_{13}$ versus warm
  (left) and cold (right) crystallinity fractions. The YSOs in Serpens
  (black dots), Taurus (red squares), Upper Sco (blue stars), and
  $\eta$ Cha (green triangles) are compared. }
\end{figure}

Figure \ref{f_f30} shows $F_{30}/F_{13}$ as a proxy for disk geometry
compared with the mean mass-averaged grain sizes and crystallinity
fractions for both components and all regions studied here. No
preferential grain size (correlation coefficient $\tau$ = -0.14, $P$ =
0.02, and $\tau$ = 0.07, $P$ = 0.33 for warm and cold components,
respectively) nor crystallinity fraction ($\tau$ = 0.09, $P$ = 0.10
for the warm, and $\tau$ = -0.19, $P$ = 0.01 for the cold component)
is apparent for any given disk geometry. Similar scatter plots result
for the mean mass-average grains sizes for only amorphous ($\tau$ =
-0.12, $P$ = 0.08 for the warm, and $\tau$ = 0.13, $P$ = 0.11 for the
cold component), or only crystalline grains ($\tau$ = 0.08, $P$ = 0.17
for the warm, and $\tau$ = -0.13, $P$ = 0.10 for the cold
component). Furthermore, no clear separation is seen between the
different regions studied. The statistically relevant samples in
Serpens and Taurus define a locus where the majority of the objects is
located in each plot, which is followed by the lower number statistics
for older regions. Figure \ref{f_f30} therefore shows not only that
grain size and crystallinity fraction are not a function of disk
geometry, but also that younger and older regions show similar
distributions of those two parameters.

\subsection{Crystallinity Fraction}
\label{s_cryst}

\begin{figure}[!h]
\begin{center}
\includegraphics[width=0.3\textwidth]{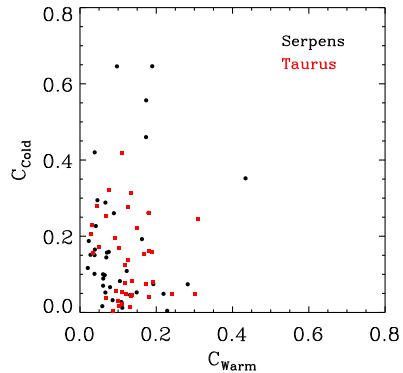}
\end{center}
\caption{\label{f_cryst} Crystalline fraction of the warm and cold
  components in Serpens (black dots) and Taurus (red squares).}
\end{figure}

The crystallinity fractions derived from the warm and cold components
($C_{\rm Warm}$ and $C_{\rm Cold}$, respectively) for Serpens and
Taurus are show in Figure \ref{f_cryst}. No strong trend of warm and
cold crystallinity fractions increasing together is seen ($\tau$ =
0.10, with $P$ = 0.10 for the entire sample). This fact implies that,
if an unique process is responsible for the crystallization of dust at
all radii, this process is not occurring at the same rate in the
innermost regions as further out in the disk. This is opposite to the
conclusion of \citet{WA09}, who derive a correlation between inner and
outer disk crystallinity from the simultaneous presence of the 11.3
and 33 $\mu$m features. The opacities of the crystalline species are
more complex than those two features alone, making the analysis here
more complete than that of \citet{WA09}. Our finding that the fraction
of crystalline material in disk surfaces varies with radius can
constrain some of the mechanisms for formation and distribution of
crystals.

\begin{figure}[!h]
\begin{center}
\includegraphics[width=0.45\textwidth]{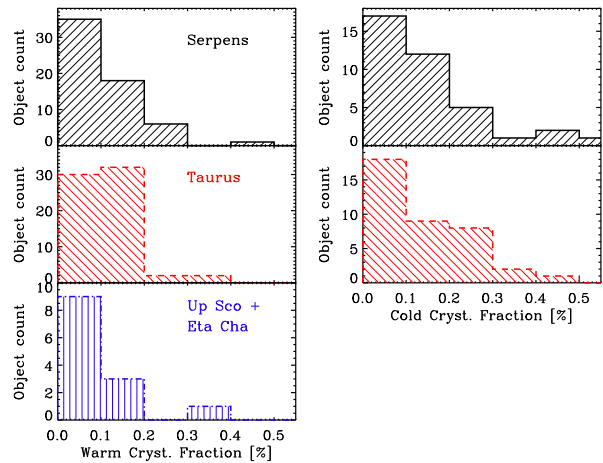}
\end{center}
\caption{\label{f_warmdist} Distribution of crystalline fractions for
  Serpens (top), Taurus (middle), and Upper Sco and $\eta$ Cha
  combined (bottom). Similar distributions and the same range of
  fractions are seen for all clusters.}
\end{figure}

\begin{figure}[!h]
\begin{center}
\includegraphics[width=0.4\textwidth]{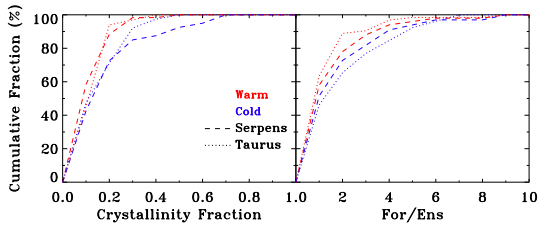}
\end{center}
\caption{\label{f_cum} Left: Cumulative fractions of the crystallinity
  fractions, for Serpens (dashed line) and Taurus (dotted
  line). Right: Cumulative fraction of the ration between the
  forsterite and enstatite fractions, for Serpens (dashed line) and
  Taurus (dotted line). The warm component is shown in red while the
  cold component is blue.}
\end{figure}

A wider spread in crystallinity fraction is observed for the cold
component than for the warm component (Figure \ref{f_warmdist}), which
is reflected in the mean crystallinity fractions for each sample
(Table \ref{t_mean}). This discrepancy could be real, or an artifact
due to the signal-to-noise ratio (S/N) being frequently lower at
longer wavelengths (cold component) than that at shorter wavelengths
(warm component), introducing a larger scatter. The difference in
Serpens is more significant ($\langle C_{\rm warm} \rangle \simeq$
11.0\% and $\langle C_{\rm cold} \rangle \simeq$ 17.5\%). The left
panel of Figure \ref{f_cum} shows the cumulative fractions as
functions of crystallinity fractions. Despite small differences
between the warm (red line) and cold (blue line) components, the two
cumulative fractions have similar behavior. If this difference is
true, there is a small fraction of T Tauri disks with a higher cold
(outer) than warm (inner) crystallinity fraction. This finding
contrasts with that derived by \citet{VB04} for the disks around 3
Herbig stars. Their spatially resolved observations infer higher
crystallinity fractions in the inner than in the outer disks, albeit
based on only 10 $\mu$m data. A larger sample of objects with good S/N
including both 10 and 20 $\mu$m data is needed to better constrain
this point. In addition, Figure \ref{f_warmdist} shows that younger
and older clusters have similar distributions of crystallinity
fractions.

\begin{figure}[!h]
\begin{center}
\includegraphics[width=0.5\textwidth]{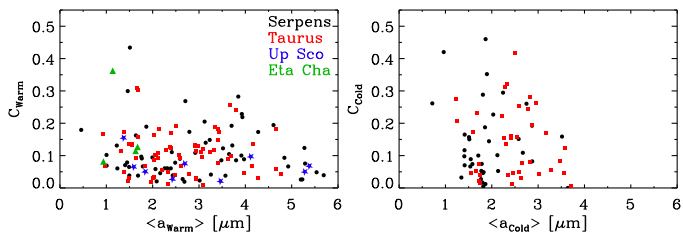}
\end{center}
\caption{\label{f_warm} Mass-averaged mean grain sizes versus the
  crystalline fraction for Serpens (black dots), Taurus (red squares),
  Upper Sco (blue stars), and $\eta$ Cha (green triangles).}
\end{figure}

As discussed by many authors, both the grain size and the degree of
crystallinity affect the silicate features, therefore it is
interesting to search for trends between these two parameters. In
Figure \ref{f_warm}, the mass-average grain sizes are compared to the
crystallinity fraction for both warm (left panel) and cold (right
panel) components. No obvious trends are seen in either component,
neither any separation between regions. This result supports the
discussion of \citet{OF10} that whatever processes govern the mean
grain size and the crystallinity in disks, they are independent from
each other.

\subsubsection{Enstatite vs. Forsterite}
\label{scomp2}

The disk models of \citet{GA04} consider chemical equilibrium of a
mixture of solid and gas at high temperatures, allowing radial mixing
of material. These models predict a predominance of forsterite in the
innermost regions of the disk, while enstatite dominates at lower
temperatures (being converted from forsterite). From the observational
point of view, data on disks around T Tauri \citep{BO08} and Herbig
Ae/Be stars \citep{JU10} have shown the opposite trend: enstatite is
more concentrated in the inner disk, while forsterite dominates the
colder, outer disk region. \citet{BO08} interpret this result as a
radial dependence of the species formation mechanisms, or a
non-equilibrium of the conditions under which the species formed,
contrary to the models assumptions.

For the regions presented in this study, it can be seen in Table
\ref{t_mean_comp} for mean cluster values and in Table \ref{t_comp}
for individual objects that the results derived from this study
generally follow those of \citet{BO08}, with more enstatite in the
warm component and, to a lesser extent, more forsterite in the cold
component. The right panel of Figure \ref{f_cum} illustrates this for
the cumulative fraction of the forsterite over enstatite ratios for
individuals disks. However, this trend is not very significant given
the uncertainties.

\subsection{The Silicate Strength-Shape Relation}
\label{ssil}

\begin{figure}[!h]
\begin{center}
\includegraphics[width=0.35\textwidth]{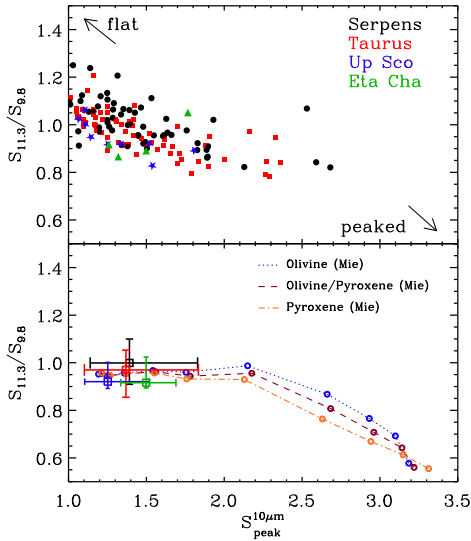}
\end{center}
\caption{\label{f_s10} Top: The ratio of normalized fluxes at 11.3 to
  9.8 $\mu$m ($S_{11.3}/S_{9.8}$) is plotted against the peak at 10
  $\mu$m ($S^{10\mu{\rm m}}_{{\rm peak}}$) for Serpens (black dots),
  Taurus (red squares), Upper Scorpius (blue stars), $\eta$
  Chamaeleontis (green triangles). Bottom: Squares show the median
  values and crosses indicate the 15 -- 85 percentile ranges of the
  distributions (top panel). Colored curves are derived from
  theoretical opacities for different mixtures by \citet{OF09}. The
  open circles correspond to different grain sizes, from left to right
  6.25, 5.2, 4.3, 3.25, 2.7, 2.0, 1.5, 1.25, 1.0 and 0.1 $\mu$m. }
\end{figure}

A correlation between the shape and the strength of the 10 $\mu$m
silicate feature from disks has been discussed extensively in the
literature \citep{VB03,KE06,OF09,PA09,OL10,OF10}. Synthetic 10 $\mu$m
features generated for different grain sizes and compositions have
been shown to fit well with observations, yielding grain size as the
important parameter responsible for such a relationship. The degree of
crystallinity of the dust also plays a role on the shape of this
feature. However, as clearly shown for EX Lup \citep{AB09}, and
supported by models \citep{MI08,OF09}, an increase in crystallinity
fraction does not change the strength of the feature, even though its
shape does change. Crystallinity is then understood as responsible for
the scatter in the strength-shape relationship, and not the
relationship itself. As a result, the strength and shape of the 10
$\mu$m silicate feature yield the typical size of the grains in the
upper layers of the disk at a few AU from the star \citep{KE07}. The
top panel of Figure \ref{f_s10} shows the results for Serpens, Taurus,
Upper Sco and $\eta$ Cha. The bottom panel presents the median values
per region, indicating the 15 -- 85 percentile ranges of the
distributions. Overlaid are the models of \citet{OF09} for different
grain sizes (0.1 -- 6.0 $\mu$m) generated for amorphous silicates of
olivine and pyroxene stoichiometry, and a 50:50 mixture. The
difference in mean ages does not correspond to a significant
difference in mean grain sizes between the different regions.

\begin{figure}[!h]
\begin{center}
\includegraphics[width=0.5\textwidth]{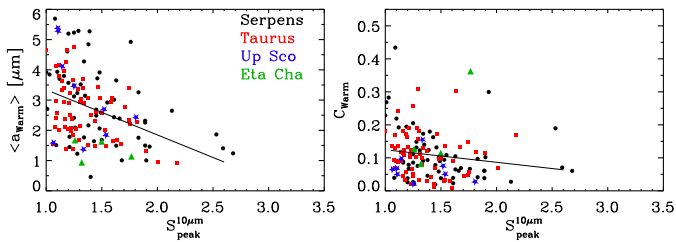}
\end{center}
\caption{\label{f_s10_2} Left panel: Strength of the 10 $\mu$m silicate
  feature ($S^{10\mu{\rm m}}_{{\rm peak}}$) versus the mass-averaged
  mean grain size for the warm component. Right panel: Strength of the
  10 $\mu$m silicate feature versus crystalline fraction for the warm
  component. The best fit relationships are shown for reference. }
\end{figure}

With the mean grain sizes derived from the spectral decomposition, it
is possible to further explore the validity of using the strength of
the 10 $\mu$m silicate feature to trace the sizes of grains in the
surfaces of disks. The left panel of Figure \ref{f_s10_2} shows the
correlation between $\langle a_{\rm warm} \rangle$ and $S^{10\mu{\rm
    m}}_{{\rm peak}}$ for all 4 samples. The Kendall $\tau$
coefficient of -0.29, $P$ = 0.01 supports the effectiveness of
$S^{10\mu{\rm m}}_{{\rm peak}}$ as a proxy for grain sizes, with
smaller values of $S^{10\mu{\rm m}}_{{\rm peak}}$ implying larger
grain sizes.

On the other hand, it is also possible to test how the degree of
crystallinity can influence the strength of the 10 $\mu$m silicate
feature. The lack of correlation between $C_{\rm warm}$ and
$S^{10\mu{\rm m}}_{{\rm peak}}$ ($\tau$ = -0.07, $P$ = 0.14), shown in
the right panel of Figure \ref{f_s10_2} for all samples, supports that
the degree of crystallinity is not the dominant parameter setting the
strength of the 10 $\mu$m silicate feature. These results argue
against the results of \citet{ST09} that find a high crystallinity
fraction and small grains fitting low strengths of the 10 $\mu$m
silicate feature. Although it may be possible to fit a few spectra
with a certain prescription, a good model should be able to explain
the robust relationship between the strength and shape of the 10
$\mu$m silicate feature observed for large numbers of disks. Despite
the many processes able to change the shape or the strength of this
feature, only grain size has so far demonstrated capability to explain
the observed trend. Our conclusion is that $S^{10\mu{\rm m}}_{{\rm
    peak}}$ and dust sizes are appropriately correlated.

\subsection{Comparison with other studies}
\label{scomp}

Dust composition results are available in the literature for the disks
in Taurus and $\eta$ Cha (see Table \ref{t_lit} for an overview).
\citet{SI09} present their analysis in $\eta$ Cha considering the same
5 dust species and three grain sizes (enstatite in their model is the
only species for which only the 2 smaller grain sizes are considered),
but for a distribution of temperatures derived using the Two Layer
Temperature Distribution (TLTD, \citealt{JU09}) decomposition
procedure. For the same 4 objects, their mean amorphous fraction is
80.1 $\pm$ 9.3 \%. This result is consistent with the 82.8 $\pm$ 12.9
\% mean amorphous fraction found here. The mean crystalline fractions
derived are 18.4 $\pm$ 10.7 \% with TLTD and 17.1 $\pm$ 12.8 \%
derived here.

\citet{ST09} present their decomposition procedure for 65 YSOs in
Taurus. This method also takes into consideration a warm and a cold
temperature, and makes use of two amorphous species (olivine and
pyroxene) with two grain sizes (small and large), and 3 crystalline
species (enstatite, forsterite and crystalline silica) of a single
size. Their mean warm amorphous fraction is 82.9 $\pm$ 19.3 \% and
warm crystalline fraction is $17.1 \pm 19.3$ \%, while here the
derived fractions are 89.0 $\pm$ 6.6 \% and 10.9 $\pm$ 6.6 \% for the
warm amorphous and crystalline fractions, respectively. For the cold
component, \citet{ST09} derive a mean cold amorphous fraction of 77.3
$\pm$ 19.9 \% and cold crystalline fraction of 22.6 $\pm$ 19.9 \%,
while here the values are 85.9 $\pm$ 10.6 \% and 13.9 $\pm$ 10.5 \%,
respectively. The consistently lower amorphous (higher crystalline)
fractions found by \citet{ST09} could be a result of their choice to
use silica in crystalline rather than amorphous form (as used
here). 

\section{Discussion}
\label{sdis}

\subsection{Dust Characteristics}
\label{sover} 

Section \ref{sres} has shown that the disk populations in the four
regions presented here, young and older, have very similar
distributions in the two main dust parameters: grain size and
composition. The large number of objects in the two young regions
studied occupy a region in parameter space of either grain size or
crystallinity fraction that is also populated by the small number of
older disks. The grain sizes derived for the cold component never
reach the biggest grain size modeled (6 $\mu$m), different from the
warm component results that span the entire range in sizes. The
crystallinity fraction does not seem to be correlated with mean grain
size, warm or cold. Whatever processes are responsible for the
crystallization of the initially amorphous grains, they should not
only be independent from the processes that govern the grain size
distribution, but they should also be able to work on bigger amorphous
grains. Alternatively, the crystalline lattice should be able to keep
itself regular during the coagulation of small crystalline dust to
create big crystalline grains.
The correlation between the strength of the 10 $\mu$m feature and the
mean grain size in disk surfaces, combined with the lack of
correlation between crystallinity fraction and $S^{10\mu{\rm m}}_{{\rm
    peak}}$, supports the wide usage of $S^{10\mu{\rm m}}_{{\rm
    peak}}$ as a proxy for dust size in literature
\citep{VB03,KE06,PA09}.

\citet{BO08} found a strong correlation between disk geometry and the
strength of the 10 $\mu$m silicate feature for a very small sample of
T Tauri stars (7 disks), which points to flatter disks having
shallower 10 $\mu$m features (i.e., big grains in the disk
surface). Using results from similar decomposition procedures,
\citet{OF10} and \citet{JU10} confirm this trend for larger samples of
T Tauri (58 disks) and Herbig Ae/Be stars (45 disks),
respectively. Those trends are much weaker than that found by
\citet{BO08}, showing a larger spread. For the current even larger
sample (139 disks), no significant trend is seen, indicating that the
earlier small sample trends may have been affected by a few
outliers. This result is similar to that found by \citet{OL10} for a
large YSO sample ($\sim$~200 objects) using the strength of the 10
$\mu$m silicate feature as a proxy for grain size (Figure 14 in that
paper). As discussed by Oliveira et al., the sedimentation models of
\citet{DD08} expect a strong correlation of larger grains in flatter
disks that is not seen. This means that sedimentation alone cannot be
responsible for the distribution of mean grain sizes in the upper
layers of protoplanetary disks around T Tauri stars. Furthermore, the
lack of correlation between crystallinity fraction and disk geometry
is not in support of the results of \citet{WA09} and \citet{ST09}, who
find a link between increasing crystallinity fraction and dust
sedimentation.

\begin{figure}[!h]
\begin{center}
\includegraphics[width=0.49\textwidth]{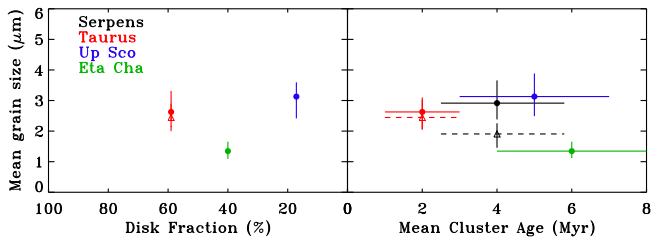}
\end{center}
\caption{\label{f_grain} Left: Mean mass-average grain sizes vs. disk
  fraction. Serpens is not included because its disk fraction is not
  yet known. Filled circles represent mean warm grain sizes, and open
  triangles represent mean cold grain sizes. Error bars for the mean
  mass-average grain sizes are estimated using a Monte Carlo approach,
  sampling the errors of the individual objects. Right: Mean grain
  sizes vs. mean cluster age. Filled circles represent results for the
  warm component, while open triangles represent the cold
  component. The black points are YSOs in Serpens, red in Taurus, blue
  in Upper Sco and green in $\eta$ Cha. }
\end{figure}

As discussed in \citet{OL10} for Serpens and Taurus, and confirmed by
the addition of considerably older samples, there is no clear
difference in the mean grain sizes in the disk surfaces with mean
cluster age, which can be seen in Figure \ref{f_grain}. This evidence
supports the discussion in that paper that the dust population
observed in the disk surface cannot be a result of a progressive,
monotonic change of state from small amorphous grains, to large, more
crystalline grains, or `grain growth and processing'. The fact that
the distribution of grain sizes in the upper layers of disks does not
change with cluster age implies that an equilibrium of the processes
of dust growth and fragmentation must exist, which also supports the
existence of small grains in disks that are millions of years old whereas
dust growth is a rapid process \citep{WE80,DD05}. That small dust is
still seen in disks in older regions like Upper Sco and $\eta$ Cha
argues that this equilibrium of processes is maintained for millions
of years, as long as the disks are optically thick, but independent of
them having a flared or flatter geometry.

\subsection{Evolution of Crystallinity with Time?}
\label{sevol}

Literature studies of disk fractions of different YSO clusters with
different mean ages show a trend of decreasing disk fraction,
i.e. disks dissipating with time, over some few millions of years
\citep{HA01,HE08}. This decrease is clearly confirmed by the lower
fraction of disks still present in the older regions studied here
(Upper Sco and $\eta$ Cha). According to current planet formation
theories, if giant planets are to be formed from gas rich disks, the
optically thin, gas-poor disks in those older regions should already
harbor (proto-)planets. Considering the evidence from small bodies in
our own Solar System that suggest considerably higher crystallinity
fractions than ISM dust (see \citealt{WO07} and \citealt{PB10} for
reviews of latest results), a crystallinity increase must occur.

\begin{figure}[!h]
\begin{center}
\includegraphics[width=0.49\textwidth]{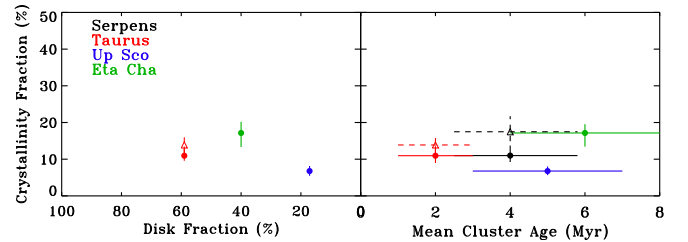}
\end{center}
\caption{\label{f_comp2} Left: Crystallinity fraction vs. disk
  fraction. Serpens is not included because its disk fraction is not
  yet known. Filled circles represent mean warm crystallinity, and
  open triangles represent mean cold crystallinity. Uncertainty for
  crystallinity fractions are estimated using a Monte Carlo approach,
  sampling the errors of the individual objects. Right: Crystallinity
  fraction vs. mean cluster age. Filled circles represent mean warm
  crystallinity, and open triangles represent mean cold
  crystallinity. The black points are YSOs in Serpens, red in Taurus,
  blue in Upper Sco and green in $\eta$ Cha. }
\end{figure}

In Figure \ref{f_comp2}, the mean crystallinity fraction per region is
plotted against two evolutionary parameters: disk fraction (left) and
mean age (right). Within the spread in individual fractions it is seen
that, just as for grain sizes, there is no strong evidence of an
increase of crystallinity fraction with either evolutionary parameter.
This implies that there is no evolution in grain sizes or
crystallinity fraction for the dust in the surface of disks over
cluster ages in the range 1 -- 8 Myr, as probed by the observations
presented here. Essentially, there is no change in these two
parameters until the disks disperse. Starting from the assumption that
initially the dust in protoplanetary disks is of ISM origin
(sub-$\mu$m in size and almost completely amorphous), it appears that
a modest level of crystallinity is established in the disk surface
early in the evolution ($\leq$ 1 Myr) and then reaches some sort of
steady state, irrespective of what is taking place in the disk
midplane. Thus, the dust in the upper layers of disks does not seem to
be a good tracer of the evolution that is taking place in the disk
interior, where dust is growing further for the formation of
planetesimals and planets, at many times higher crystallinity
fractions, to be consistent with evidence from Solar system bodies.

If this is the case, within 1 Myr this surface dust must be
crystallized to the observed fraction ($\sim$10 -- 20~\%). This result
puts constraints on the formation of circumstellar disks. One
possibility is that this crystallization of the dust in disks mostly
occurs during the embedded phase. In this early stage of star
formation, where large quantities of material are still accreting
towards the protostar, a fraction of the infalling material comes very
close to the protostar and is heated to temperatures $>$800 K before
it moves outwards in the disk. Alternatively, accretion shocks or
episodic heating events could be responsible for thermally annealing
the dust in the disk surface.

The 2-D models of \citet{VD10} treat the radial evolution of crystals
in time. According to these models, 100~\% of the dust in the inner
disk ($\leq$ 1 AU) is crystallized within 1 Myr. With time, the inner
disk crystalline fraction drops as the disk spreads, and crystalline
material is transported to outer parts of the disk. These models can
help explain the rapid crystallization required to account for our
results. However, the models do show a decrease in inner disk ($\leq$
1 AU) crystallinity fraction with time, which is not supported by our
results. Since these models do not discriminate on vertical structure,
but rather present crystallinity fractions that are integrated over
all heights at a given radius, this decrease in crystallinity fraction
is not necessarily connected to the surface of the disk. Thus the
decrease in crystallinity fraction with time found in the models of
\citet{VD10} could be explained as a decrease in crystallinity
fraction just in the disk midplane where the bulk of the mass resides,
but not in the surface layers, as our data indicate. That would imply
that radial mixing of these crystals is more efficient than vertical
mixing, which is responsible for the crystallinity fraction decrease
in the disk midplane.

According to the models of \citet{CI07} for outward transport of high
temperature materials, variations in radial transport dynamics with
height produce vertical gradients in the crystalline fractions, such
that the upper layers of the disk will have lower crystallinity
fractions than the midplane population. If that is the case, the
observations discussed here, which probe the disk surface only, lead
to lower limits on the real crystalline fraction of disk midplanes. In
this scenario, planets (and comets) forming in the disk midplane would
have higher crystalline abundances than those derived here for the
disk surfaces, which are compatible with what has been observed in our
Solar System. However, this model does not make predictions for the
time evolution of the systems. Combining the vertical and radial
mixing processes with evolutionary models such as those of
\citet{VD10} are needed to investigate whether older and younger disks
could still show the same distribution of crystallinity fractions in
the upper layers of disks, as observed here.

\section{Conclusions}
\label{scon}

This paper presents the spectral decomposition of Spitzer/IRS spectra
using the B2C decomposition model of \citet{OF10}. Mineralogical
compositions and size distributions of dust grains in the surface
layers of protoplanetary disks are derived for 139 YSOs belonging to
four young star clusters using the same method.

Serpens and Taurus are used as prototypes of young regions, where most
stars are still surrounded by disks, while Upper Sco and $\eta$ Cha
represent the older bin of disk evolution, where a large fraction of
the disks have already dissipated but some massive protoplanetary
disks are left. The large number of objects analyzed allows
statistical results that point to the main processes that affect the
grain size distribution and composition of dust in protoplanetary
disks. Furthermore, the usage of the same analysis method for regions
of different mean ages allow a study of evolution of the dust
parameters with time.

Our large sample does not show a preferential grain size or
crystallinity fraction with disk geometry, contrary to earlier
analyses based on smaller samples. Also, younger and older regions
have very similar distributions. The difference between mean
mass-averaged grain sizes for the warm and cold components of a given
star-forming region is small, however a considerable difference is
seen between the ranges of grain sizes spanned in both components. The
cold mass-averaged grain sizes never reach the biggest size modelled
(6 $\mu$m) while the warm mass-averaged grain sizes span the entire
range of sizes modeled. The crystallinity fractions derived for inner
(warm) and outer (cold) disks are typically 10 -- 20\%{}, and not
correlated. The cold crystallinity fraction shows a larger spread than
the warm. No strong difference is seen between the overall mean warm
and cold crystallinity fraction. Within the crystalline dust
population, more enstatite is found in the warm component and more
forsterite in the cold component. The differences are not very
significant, however.

The results of the spectral decomposition support the usage of the
strength of the 10 $\mu$m silicate feature ($S^{10\mu{\rm m}}_{{\rm
    peak}}$) as a proxy of the mean grain size of dust in the disk
surface. This is supported by the correlation between $S^{10\mu{\rm
    m}}_{{\rm peak}}$ and mean grain size and lack of correlation with
the mean crystallinity.

Mean cluster ages and disk fractions are used as indicators of the
evolutionary stage of the different populations. Our results show that
the different regions have similar distributions of mean grain sizes
and crystallinity fractions regardless of the spread in mean ages of 1
-- 8 Myr. Thus, despite the fact that the majority of disks dissipate
within a few Myr, the surface dust properties do not depend on age for
those disks that have not yet dissipated in the 1 -- 8 Myr range. This
points to a rapid change in the composition and crystallinity of the
dust in the early stages ($\leq$ 1 Myr) that is maintained essentially
until the disks dissipate.

\acknowledgements Astrochemistry at Leiden is supported by a Spinoza
grant from the Netherlands Organization for Scientific Research (NWO)
and by the Netherlands Research School for Astronomy (NOVA)
grants. This work is based on observations made with the Spitzer Space
Telescope, which is operated by the Jet Propulsion Laboratory,
California Institute of Technology under a contract with NASA. Support
for this work was provided by NASA through an award issued by
JPL/Caltech.

\begin{table*}
\centering
\caption{Characteristics of the star-forming regions presented in
  this work.}
\label{t_overview}
\begin{tabular}{l | c c c c}
\hline
Region & Dist (pc) & Mean Age (Myr) & Disk Fraction & Ref. \\
\hline
Serpens    & 259--415 & 2--6    & --         & 1, 2      \\
Taurus     & 140      & $\sim$2 & $\sim$60\% & 3, 4, 5   \\
Up Sco     & 145      & $\sim$5 & $\sim$17\% & 6, 7, 8   \\
$\eta$ Cha &  97      & $\sim$6 & $\sim$40\% & 9, 10, 11 \\
\hline
\end{tabular}

References: 1: The distance to Serpens is still under debate,
different methods yield distances ranging from 259 \citep{ST96} to 415
pc \citep{DZ10}; 2: From \citet{OL09}, using $d$ = 259 pc; 3: From
\citet{KE94}; 4: From \citet{LH01}; 5: From \citet{LU10}; 6: From
\citet{DZ99}; 7: From \citet{BL78}; 8: From \citet{CA06}; 9: From
\citet{MA99}; 10: From \citet{LS04}; 11: From \citet{ME05}.

\end{table*}

\begin{table*}
\centering
\caption{B2C mean composition of each star-forming region.}
\label{t_mean_comp}
\begin{tabular}{l | c c c c}
\hline
Region & Oli/Pyr$^a$ & Silica & Forsterite & Enstatite  \\
\hline
\multicolumn{5}{c}{Warm Component} \\
Serpens    & 81.3$\pm$11.7\% &  7.8$\pm$6.5\% &  5.8$\pm$4.9\% &  5.2$\pm$4.9\% \\
Taurus     & 79.4$\pm$9.4\%  &  9.6$\pm$7.0\% &  4.4$\pm$3.3\% &  6.5$\pm$4.5\% \\
Up Sco     & 89.7$\pm$4.7\%  &  3.5$\pm$3.4\% &  2.6$\pm$1.1\% &  4.1$\pm$3.1\% \\
$\eta$ Cha & 75.0$\pm$13.4\% &  7.8$\pm$4.6\% &  6.8$\pm$4.0\% & 10.3$\pm$9.8\% \\
\multicolumn{5}{c}{Cold Component} \\
Serpens    & 68.0$\pm$20.1\% & 14.4$\pm$12.3\% &  9.5$\pm$9.5\% &  8.0$\pm$8.0\% \\
Taurus     & 64.7$\pm$14.3\% & 21.3$\pm$11.4\% &  8.6$\pm$8.6\% &  5.3$\pm$5.3\% \\
\hline
\end{tabular}

$^a$ Amorphous olivine and pyroxene combined. 
\end{table*}

\begin{table*}
\centering
\caption{B2C mean grain size and crystallinity parameters for each
  star-forming region.}
\begin{minipage}{140mm}
\label{t_mean}
\begin{tabular}{l | c c c c c c c}
\hline
Region & Number & $\langle a_{\rm warm} \rangle$ & $\langle a_{\rm cold} \rangle$ & $\langle \alpha_{\rm warm} \rangle$ & $\langle \alpha_{\rm cold} \rangle$ & $\langle C_{\rm warm} \rangle$ & $\langle C_{\rm cold} \rangle$ \\
       &        & $\mu$m      & $\mu$m   &   &   & \%  & \% \\
\hline
Serpens    & 60 & 2.9$\pm$1.3 & 1.9$\pm$0.6 & -2.75$\pm$0.39 & -3.16$\pm$0.18 & 11.0$\pm$6.9 & 17.5$\pm$12.4 \\
Taurus     & 66 & 2.6$\pm$0.9 & 2.4$\pm$0.6 & -2.83$\pm$0.31 & -3.02$\pm$0.15 & 10.9$\pm$5.6 & 13.9$\pm$10.1 \\
Up Sco     &  9 & 3.1$\pm$1.5 &  --         & -3.33$\pm$0.18 & --             &  6.8$\pm$3.3 & --        \\
$\eta$ Cha &  4 & 1.3$\pm$0.4 &  --         & -2.71$\pm$0.39 & --             & 17.1$\pm$10.6 & --        \\
\hline
\end{tabular}
\end{minipage}
\end{table*}

\begin{table*}
\centering
\caption{Comparison of mean mineralogical results from this study with
  literature studies.}
\label{t_lit}
\begin{tabular}{l | c c | c c}
\hline
Region & \multicolumn{2}{c}{This work} & \multicolumn{2}{c}{Literature} \\
       & Amorphous & Crystalline & Amorphous & Crystalline \\
\hline
\multicolumn{5}{c}{Warm Component} \\
Taurus     & 89.0$\pm$6.6 \%  & 10.9$\pm$6.6 \%  & 82.9$\pm$19.3 \% & 17.1$\pm$19.3 \% $^a$ \\
$\eta$ Cha & 82.8$\pm$12.9 \% & 17.1$\pm$12.8 \% & 80.1$\pm$9.3 \%  & 18.4$\pm$10.7 \% $^b$ \\
\multicolumn{5}{c}{Cold Component} \\
Taurus     & 85.9$\pm$10.6 \% & 13.9$\pm$10.5 \% & 77.3$\pm$19.9 \% & 22.6$\pm$19.9 \% $^a$ \\
\hline
\end{tabular}

$^a$ \citet{ST09}; $^b$ \citet{SI09}   
\end{table*}

\clearpage

\appendix

\section{Relative Abundances of Species}
\label{sabun}

The relative abundances, as resulting from the B2C compositional
fitting (\S~\ref{s_spdecomp}) to the IRS spectra of protoplanetary
disks in Serpens, Taurus, Upper Sco and $\eta$ Cha are shown in Table
\ref{t_comp}. Since the opacities of amorphous olivine and pyroxene
are degenerate, the abundances of these two species have been added
into one, marked `Oli/Pyr' in the Table. Furthermore, `Sil' designates
the amorphous silica, and the crystalline enstatite and forsterite are
marked as `Ens' and `For', respectively. In the table, the first line
of a given object corresponds to the results of the fit to the warm
component and the second line to the results of the cold component.
For some objects (20 in Serpens, 28 in Taurus, all in Upper Sco and
$\eta$ Cha), the S/N drops considerably at longer wavelengths and the
results of the procedure are no longer reliable. For these sources,
the cold component could not be fitted satisfactorily and, in Table
\ref{t_comp}, only the warm component results are shown.

\clearpage

%

\begin{table}
\centering
\tiny
\caption{Dust composition derived using the ``B2C''
  procedure.$^b$} 
\label{t_comp}
\begin{tabular}{lc|cccc|cccc|cc}
\hline
ID & SpT & Oli/Pyr$^a$ \% & Ens \% & For \% & Sil \% & Oli/Pyr \% & Ens \% & For \% & Sil \% & Oli/Pyr \% & Sil \% \\
   &     & (0.1 $\mu$m) & (0.1 $\mu$m) & (0.1 $\mu$m) & (0.1 $\mu$m) & (1.5 $\mu$m) & (1.5 $\mu$m) & (1.5 $\mu$m) & (1.5 $\mu$m) & (6.0 $\mu$m) & (6.0 $\mu$m) \\
\hline
\multicolumn{12}{c}{Serpens} \\
1        & K2   & 29.7$_{-15.3}^{+10.7}$ & 1.0$_{-0.4}^{+2.0}$   & 2.6$_{-1.4}^{+2.6}$    & 0.1$_{-0.0}^{+0.9}$    & 53.7$_{-23.0}^{+16.5}$ & 3.5$_{-2.4}^{+3.5}$   & 0.0$_{-0.0}^{+2.9}$    & 1.0$_{-0.0}^{+2.1}$    & 7.8$_{-0.0}^{+3.8}$    & 0.6$_{-0.0}^{+1.9}$        \\
$\cdots$ &      & 38.4$_{-15.8}^{+14.5}$ & 0.0$_{-0.0}^{+1.6}$   & 8.7$_{-3.6}^{+5.5}$    & 1.2$_{-0.0}^{+3.0}$    & 29.3$_{-15.3}^{+6.2}$  & 6.8$_{-2.8}^{+4.3}$   & 0.2$_{-0.0}^{+1.1}$    & 0.3$_{-0.0}^{+1.4}$    & 9.7$_{-4.4}^{+4.7}$    & 5.5$_{-1.6}^{+1.4}$        \\
3        & M0   & 26.5$_{-7.3}^{+15.9}$  & 0.0$_{-0.0}^{+2.4}$   & 0.9$_{-0.4}^{+3.5}$    & 0.0$_{-0.0}^{+1.7}$    & 43.9$_{-18.6}^{+19.8}$ & 0.2$_{-0.2}^{+4.1}$   & 2.7$_{-2.1}^{+3.8}$    & 0.0$_{-0.0}^{+2.9}$    & 15.7$_{-2.8}^{+3.2}$   & 0.0$_{-0.0}^{+1.0}$        \\
$\cdots$ &      & 27.3$_{-8.6}^{+33.7}$  & 0.0$_{-0.0}^{+10.5}$  & 10.2$_{-4.9}^{+19.1}$  & 0.0$_{-4.0}^{+16.3}$   & 32.9$_{-9.9}^{+34.1}$  & 4.9$_{-5.5}^{+19.2}$  & 0.0$_{-2.9}^{+13.9}$   & 12.7$_{-5.6}^{+22.5}$  & 10.2$_{-4.7}^{+7.4}$   & 1.8$_{-0.6}^{+2.2}$        \\
6        & K5   & 25.4$_{-6.0}^{+4.5}$   & 0.0$_{-0.0}^{+0.7}$   & 0.8$_{-0.0}^{+1.1}$    & 0.0$_{-0.0}^{+0.3}$    & 38.2$_{-7.4}^{+11.6}$  & 1.0$_{-0.0}^{+1.4}$   & 1.0$_{-0.0}^{+1.4}$    & 0.0$_{-0.0}^{+1.4}$    & 28.8$_{-5.0}^{+4.5}$   & 4.8$_{-1.8}^{+3.6}$        \\
$\cdots$ &      & 6.4$_{-3.3}^{+3.8}$    & 4.1$_{-2.3}^{+2.1}$   & 5.1$_{-0.0}^{+3.4}$    & 0.5$_{-0.0}^{+1.1}$    & 46.1$_{-20.3}^{+19.8}$ & 4.7$_{-0.0}^{+6.2}$   & 1.2$_{-0.0}^{+2.3}$    & 2.7$_{-0.0}^{+4.4}$    & 25.4$_{-7.6}^{+8.6}$   & 3.7$_{-0.0}^{+2.2}$        \\
7        & M0   & 3.7$_{-0.9}^{+3.1}$    & 0.0$_{-0.0}^{+2.9}$   & 0.8$_{-0.0}^{+4.3}$    & 0.0$_{-0.0}^{+0.9}$    & 75.6$_{-23.9}^{+26.0}$ & 0.0$_{-0.1}^{+5.6}$   & 3.7$_{-3.0}^{+4.7}$    & 0.0$_{-0.0}^{+5.1}$    & 0.0$_{-0.0}^{+9.8}$    & 16.1$_{-13.7}^{+4.4}$      \\
$\cdots$ &      & 0.0$_{-0.5}^{+3.4}$    & 7.2$_{-6.9}^{+4.9}$   & 8.1$_{-6.9}^{+5.3}$    & 14.1$_{-11.5}^{+8.3}$  & 6.0$_{-6.0}^{+7.7}$    & 14.1$_{-11.7}^{+8.8}$ & 0.0$_{-0.1}^{+3.0}$    & 24.7$_{-19.7}^{+14.3}$ & 25.6$_{-13.9}^{+14.0}$ & 0.0$_{-2.6}^{+3.5}$        \\
9        & --   & 58.3$_{-8.8}^{+10.9}$  & 0.0$_{-0.0}^{+1.4}$   & 0.5$_{-0.0}^{+1.1}$    & 0.0$_{-0.0}^{+0.7}$    & 27.4$_{-7.2}^{+5.7}$   & 3.8$_{-2.1}^{+2.4}$   & 2.5$_{-2.0}^{+1.5}$    & 0.0$_{-0.0}^{+1.3}$    & 3.6$_{-0.0}^{+4.1}$    & 3.7$_{-0.0}^{+1.7}$        \\
$\cdots$ &      & 24.5$_{-14.5}^{+16.3}$ & 0.0$_{-0.0}^{+3.5}$   & 14.4$_{-9.8}^{+13.9}$  & 16.9$_{-15.0}^{+24.0}$ & 9.4$_{-6.1}^{+6.7}$    & 0.0$_{-0.8}^{+3.8}$   & 0.0$_{-0.0}^{+1.5}$    & 7.5$_{-5.2}^{+5.8}$    & 24.4$_{-14.3}^{+16.2}$ & 2.7$_{-3.2}^{+7.2}$        \\
10       & --   & 1.9$_{-0.6}^{+5.9}$    & 0.0$_{-0.0}^{+2.5}$   & 2.3$_{-0.7}^{+2.1}$    & 4.3$_{-1.4}^{+2.4}$    & 37.6$_{-16.8}^{+17.3}$ & 7.2$_{-3.5}^{+4.2}$   & 5.3$_{-2.9}^{+6.6}$    & 0.2$_{-0.0}^{+4.0}$    & 31.6$_{-8.3}^{+18.5}$  & 9.5$_{-4.4}^{+7.6}$        \\
$\cdots$ &      & 37.4$_{-9.4}^{+29.1}$  & 2.0$_{-0.0}^{+10.1}$  & 0.0$_{-0.0}^{+9.2}$    & 0.0$_{-0.0}^{+19.4}$   & 39.0$_{-10.3}^{+23.6}$ & 0.0$_{-0.0}^{+20.1}$  & 3.3$_{-1.3}^{+9.6}$    & 0.0$_{-0.0}^{+9.3}$    & 14.8$_{-1.2}^{+35.1}$  & 3.6$_{-0.5}^{+14.1}$       \\
14       & M2   & 11.1$_{-0.0}^{+46.8}$  & 0.0$_{-0.0}^{+6.0}$   & 3.0$_{-0.0}^{+7.1}$    & 0.0$_{-0.0}^{+6.9}$    & 54.7$_{-20.3}^{+31.3}$ & 4.8$_{-0.0}^{+20.8}$  & 0.0$_{-0.0}^{+4.9}$    & 0.0$_{-0.0}^{+8.7}$    & 13.5$_{-5.8}^{+9.3}$   & 12.8$_{-4.0}^{+10.4}$      \\
15       & --   & 35.6$_{-7.0}^{+12.5}$  & 0.2$_{-0.0}^{+1.5}$   & 2.9$_{-1.3}^{+1.6}$    & 0.0$_{-0.0}^{+0.6}$    & 51.3$_{-16.1}^{+9.3}$  & 3.0$_{-1.5}^{+3.4}$   & 0.0$_{-0.0}^{+1.6}$    & 0.8$_{-0.0}^{+1.8}$    & 1.2$_{-0.0}^{+2.2}$    & 5.0$_{-0.5}^{+2.0}$        \\
$\cdots$ &      & 36.3$_{-12.9}^{+10.4}$ & 2.4$_{-1.6}^{+1.8}$   & 2.2$_{-1.0}^{+3.5}$    & 1.1$_{-0.9}^{+2.8}$    & 34.9$_{-12.2}^{+9.9}$  & 0.0$_{-0.0}^{+1.0}$   & 5.4$_{-2.1}^{+2.0}$    & 0.0$_{-0.6}^{+2.1}$    & 14.6$_{-4.6}^{+4.7}$   & 3.0$_{-1.9}^{+1.4}$        \\
21       & --   & 2.2$_{-0.0}^{+5.0}$    & 0.0$_{-0.0}^{+1.6}$   & 2.7$_{-0.0}^{+3.7}$    & 0.1$_{-0.0}^{+3.6}$    & 20.2$_{-12.9}^{+10.1}$ & 0.1$_{-0.0}^{+4.2}$   & 19.2$_{-9.1}^{+11.1}$  & 0.0$_{-0.0}^{+4.2}$    & 43.3$_{-21.3}^{+.6.5}$ & 12.3$_{-9.1}^{+8.6}$       \\
$\cdots$ &      & 33.0$_{-20.7}^{+18.1}$ & 1.6$_{-0.0}^{+5.2}$   & 0.0$_{-0.0}^{+2.3}$    & 1.3$_{-0.0}^{+11.9}$   & 42.5$_{-24.9}^{+22.6}$ & 2.7$_{-0.0}^{+14.8}$  & 0.5$_{-0.0}^{+3.8}$    & 0.0$_{-0.0}^{+14.2}$   & 9.6$_{-3.3}^{+8.7}$    & 8.7$_{-7.5}^{+5.2}$        \\
29       & M2   & 1.4$_{-0.0}^{+6.4}$    & 0.0$_{-0.0}^{+3.1}$   & 1.6$_{-0.0}^{+7.8}$    & 2.7$_{-0.3}^{+6.6}$    & 71.0$_{-49.2}^{+26.0}$ & 0.0$_{-0.0}^{+ 9.3}$  & 4.4$_{-3.1}^{+9.4}$    & 0.0$_{-0.0}^{+7.9}$    & 13.0$_{-7.}^{+6.4}$    & 5.6$_{-3.1}^{+14.7}$       \\
30       & M1   & 0.0$_{-0.0}^{+9.7}$    & 0.1$_{-0.0}^{+4.4}$   & 3.4$_{-1.2}^{+4.6}$    & 4.5$_{-0.2}^{+6.8}$    & 12.5$_{-6.4}^{+8.5}$   & 12.5$_{-3.4}^{+8.1}$  & 1.3$_{-0.9}^{+5.9}$    & 1.5$_{-0.8}^{+5.0}$    & 64.2$_{-47.0}^{+18.1}$ & 0.0$_{-0.0}^{+21.4}$       \\
$\cdots$ &      & 10.3$_{-13.8}^{+51.7}$ & 1.7$_{-3.2}^{+15.4}$  & 5.5$_{-4.3}^{+14.1}$   & 8.1$_{-6.3}^{+21.6}$   & 8.0$_{-19.4}^{+58.7}$  & 9.1$_{-5.1}^{+14.9}$  & 29.6$_{-24.3}^{+44.9}$ & 11.4$_{-10.3}^{+37.5}$ & 14.0$_{-10.3}^{+36.3}$ & 2.1$_{-2.3}^{+9.6}$        \\
31       & --   & 1.2$_{-0.0}^{+0.8}$    & 0.0$_{-0.0}^{+0.8}$   & 0.8$_{-0.0}^{+0.5}$    & 0.0$_{-0.0}^{+1.0}$    & 15.0$_{-1.5}^{+2.6}$   & 2.5$_{-0.0}^{+1.2}$   & 3.4$_{-1.3}^{+1.5}$    & 0.4$_{-0.0}^{+0.4}$    & 76.1$_{-5.4}^{+7.8}$   & 0.6$_{-0.0}^{+1.0}$        \\
$\cdots$ &      & 21.8$_{-4.9}^{+5.3}$   & 2.4$_{-0.0}^{+1.6}$   & 0.0$_{-0.0}^{+0.7}$    & 16.7$_{-7.9}^{+8.7}$   & 24.0$_{-4.0}^{+4.7}$   & 2.3$_{-0.0}^{+2.0}$   & 0.5$_{-0.0}^{+1.7}$    & 5.4$_{-2.6}^{+4.4}$    & 9.2$_{-5.2}^{+4.8}$    & 17.6$_{-4.7}^{+3.3}$       \\
36       & K5   & 22.3$_{-6.2}^{+5.2}$   & 0.9$_{-0.0}^{+1.6}$   & 0.6$_{-0.0}^{+1.3}$    & 0.0$_{-0.0}^{+0.8}$    & 51.9$_{-12.3}^{+12.1}$ & 1.0$_{-0.6}^{+2.9}$   & 3.3$_{-2.5}^{+2.8}$    & 0.0$_{-0.0}^{+1.6}$    & 10.8$_{-2.6}^{+2.8}$   & 9.0$_{-4.8}^{+2.7}$        \\
$\cdots$ &      & 37.4$_{-15.4}^{+11.0}$ & 0.0$_{-0.0}^{+3.9}$   & 1.6$_{-1.0}^{+2.7}$    & 2.9$_{-2.5}^{+8.6}$    & 38.1$_{-14.3}^{+11.0}$ & 0.0$_{-0.0}^{+2.4}$   & 0.0$_{-0.0}^{+2.9}$    & 0.2$_{-0.0}^{+8.7}$    & 19.6$_{-7.7}^{+5.9}$   & 0.1$_{-1.4}^{+0.3}$        \\
40       & M7   & 0.0$_{-0.0}^{+8.3}$    & 0.0$_{-0.0}^{+8.4}$   & 2.9$_{-0.8}^{+4.4}$    & 10.9$_{-4.1}^{+4.6}$   & 41.2$_{-25.0}^{+17.3}$ & 22.5$_{-8.8}^{+11.0}$ & 17.9$_{-7.4}^{+11.6}$  & 0.0$_{-0.0}^{+5.5}$    & 4.5$_{-0.2}^{+9.7}$    & 0.0$_{-0.0}^{+9.8}$        \\
$\cdots$ &      & 24.2$_{-18.3}^{+15.5}$ & 16.6$_{-8.4}^{+11.2}$ & 3.1$_{-0.0}^{+16.4}$   & 0.0$_{-0.0}^{+4.8}$    & 18.2$_{-13.8}^{+9.5}$  & 2.1$_{-1.9}^{+12.8}$  & 13.4$_{-9.8}^{+7.8}$   & 0.0$_{-0.0}^{+3.0}$    & 0.2$_{-0.0}^{+9.7}$    & 22.3$_{-12.6}^{+11.2}$     \\
41       & K2   & 0.0$_{-0.0}^{+18.9}$   & 6.3$_{-0.0}^{+19.1}$   & 3.9$_{-0.0}^{+16.0}$  & 0.8$_{-0.0}^{+8.6}$    & 66.6$_{-58.2}^{+6.8}$  & 3.4$_{-0.0}^{+16.3}$   & 16.4$_{-8.6}^{+11.3}$ & 0.0$_{-0.0}^{+6.3}$    & 2.6$_{-0.0}^{+13.8}$   & 0.0$_{-0.0}^{+14.8}$       \\
43       & M0.5 & 0.0$_{-0.0}^{+13.0}$   & 0.0$_{-0.0 }^{+4.0}$  & 4.2$_{-0.0}^{+5.7}$    & 4.6$_{-0.0}^{+6.1}$    & 42.1$_{-27.0}^{+19.1}$ & 7.1$_{-5.1}^{+8.9}$   & 2.7$_{-2.0 }^{+7.1}$   & 0.0$_{-0.0}^{+3.5}$    & 37.2$_{-19.3}^{+15.8}$ & 2.0$_{-0.0}^{+12.7}$       \\
48       & M5.5 & 7.0$_{-0.1}^{+17.5}$   & 0.1$_{-0.0}^{+8.4}$   & 2.0$_{-0.0}^{+10.2}$   & 0.6$_{-0.6}^{+3.0}$    & 73.3$_{-35.2}^{+28.2}$ & 0.0$_{-0.0}^{+9.2}$   & 0.0$_{-0.1}^{+13.8}$   & 1.1$_{-0.7}^{+9.9}$    & 6.0$_{-2.1}^{+8.4}$    & 9.9$_{-6.9}^{+4.6}$        \\
$\cdots$ &      & 46.0$_{-40.6}^{+23.9}$ & 5.1$_{-3.3}^{+10.9}$  & 2.5$_{-1.1}^{+10.5}$   & 0.3$_{-2.3}^{+21.3}$   & 18.7$_{-17.6}^{+10.0}$ & 0.0$_{-0.1}^{+7.5}$   & 4.1$_{-3.5}^{+12.0}$   & 8.6$_{-8.1}^{+5.8}$    & 7.4$_{-4.7}^{+5.6}$    & 7.5$_{-6.6}^{+4.4}$        \\
53       & M2.5 & 0.0$_{-0.0}^{+36.7}$   & 5.0$_{-0.0}^{+26.6}$  & 5.4$_{-0.0}^{+18.5}$   & 5.9$_{-0.0}^{+20.1}$   & 30.2$_{-14.6}^{+18.2}$ & 10.1$_{-5.8}^{+19.9}$ & 0.0$_{-0.0}^{+11.6}$   & 2.3$_{-0.7}^{+15.6}$   & 32.9$_{-6.7}^{+27.1}$  & 8.1$_{-5.3}^{+9.4}$        \\
55       & K2   & 16.9$_{-4.0}^{+11.3}$  & 1.6$_{-0.0}^{+2.7}$   & 2.3$_{-0.9}^{+2.5}$    & 0.0$_{-0.0}^{+1.1}$    & 71.0$_{-26.2}^{+16.2}$ & 0.0$_{-0.0}^{+3.0}$   & 0.0$_{-0.0}^{+2.7}$    & 1.9$_{-0.0}^{+3.0}$    & 4.2$_{-0.8}^{+1.5}$    & 2.1$_{-0.7}^{+0.9}$        \\
$\cdots$ &      & 27.7$_{-23.1}^{+81.3}$ & 4.8$_{-1.1}^{+36.6}$  & 21.5$_{-14.2}^{+34.7}$ & 0.0$_{-0.0}^{+22.9}$   & 20.5$_{-20.0}^{+71.1}$ & 15.7$_{-9.9}^{+72.9}$ & 0.0$_{-0.1}^{+22.3}$   & 4.8$_{-4.2}^{+65.6}$   & 2.9$_{-1.6}^{+5.8}$    & 1.9$_{-1.8}^{+6.0}$        \\
56       & --   & 7.8$_{-2.0}^{+8.3}$    & 1.7$_{-0.8}^{+1.9}$   & 2.5$_{-0.6}^{+2.4}$    & 1.1$_{-0.0}^{+3.4}$    & 44.6$_{-28.5}^{+16.6}$ & 5.0$_{-2.7}^{+7.9}$   & 1.7$_{-0.5}^{+7.3}$    & 0.0$_{-0.0}^{+4.6}$    & 29.5$_{-11.8}^{+11.8}$ & 6.1$_{-0.1}^{+12.2}$       \\
$\cdots$ &      & 44.4$_{-17.0}^{+22.2}$ & 1.6$_{-1.0}^{+5.1}$   & 0.0$_{-0.0}^{+6.4}$    & 2.2$_{-0.0}^{+21.1}$   & 29.8$_{-12.0}^{+11.0}$ & 0.0$_{-0.0}^{+15.8}$  & 1.1$_{-0.0}^{+7.5}$    & 0.0$_{-0.0}^{+3.2}$    & 19.3$_{-7.9}^{+24.2}$  & 1.6$_{-0.0}^{+27.0}$       \\
57       & --   & 2.2$_{-0.0}^{+22.8}$   & 0.0$_{-0.0}^{+9.1}$   & 5.7$_{-1.1}^{+7.2}$    & 3.0$_{-1.4}^{+3.2}$    & 72.0$_{-40.9}^{+29.0}$ & 10.5$_{-6.6}^{+13.3}$ & 0.0$_{-0.0}^{+7.1}$    & 0.0$_{-0.0}^{+7.1}$    & 6.5$_{-0.0}^{+19.1}$   & 0.0$_{-0.0}^{+9.5}$        \\
$\cdots$ &      & 23.0$_{-5.8}^{+76.9}$  & 0.0$_{-0.0}^{+79.5}$  & 14.4$_{-10.4}^{+46.6}$ & 0.7$_{-0.0}^{+131.2}$  & 13.1$_{-4.8}^{+55.6}$  & 0.3$_{-0.0}^{+27.7}$  & 4.5$_{-0.0}^{+61.3}$   & 1.7$_{-4.5}^{+25.5}$   & 25.1$_{-3.7}^{+85.7}$  & 17.2$_{-5.0}^{+81.4}$      \\
58       & K7   & 6.6$_{-0.5}^{+3.2}$    & 0.0$_{-0.0}^{+1.0}$   & 0.6$_{-0.0}^{+1.1}$    & 1.2$_{-0.0}^{+0.7}$    & 29.9$_{-6.7}^{+4.6}$   & 1.0$_{-0.7}^{+2.2}$   & 7.1$_{-2.1}^{+3.0}$    & 0.0$_{-0.0}^{+1.8}$    & 28.3$_{-11.5}^{+6.5}$  & 25.4$_{-8.7}^{+6.4}$       \\
$\cdots$ &      & 32.2$_{-16.4}^{+16.8}$ & 3.2$_{-0.0}^{+15.6}$  & 0.0$_{-0.0}^{+3.7}$    & 0.0$_{-0.0}^{+7.2}$    & 57.0$_{-36.3}^{+24.9}$ & 0.0$_{-0.0}^{+7.5}$   & 0.0$_{-0.0}^{+8.0}$    & 0.0$_{-0.0}^{+6.2}$    & 4.7$_{-1.4}^{+2.9}$    & 2.8$_{-0.0}^{+2.5}$        \\
60       & M0.5 & 0.0$_{-0.0}^{+10.1}$   & 9.8$_{-0.0}^{+23.2}$  & 5.2$_{-0.0}^{+11.8}$   & 0.0$_{-0.0}^{+3.9}$    & 41.7$_{-21.7}^{+31.4}$ & 8.5$_{-5.2}^{+13.1}$  & 3.3$_{-2.3}^{+14.0}$   & 0.0$_{-0.0}^{+13.7}$   & 4.9$_{-1.7}^{+14.8}$   & 26.5$_{-14.6}^{+17.0}$     \\
61       & M0   & 12.2$_{-0.0}^{+7.2}$   & 0.0$_{-0.0}^{+0.6}$   & 1.8$_{-0.0}^{+1.3}$    & 0.0$_{-0.0}^{+1.1}$    & 52.1$_{-13.2}^{+11.9}$ & 3.1$_{-0.0}^{+5.1}$   & 6.2$_{-3.0}^{+3.6}$    & 0.0$_{-0.0}^{+2.4}$    & 11.3$_{-3.0}^{+4.2}$   & 13.3$_{-6.3}^{+1.8}$       \\
$\cdots$ &      & 38.2$_{-7.6}^{+3.7}$   & 0.0$_{-0.0}^{+0.8}$   & 0.1$_{-0.0}^{+0.9}$    & 0.6$_{-0.0}^{+1.7}$    & 38.7$_{-6.3}^{+6.6}$   & 0.0$_{-0.0}^{+1.6}$   & 1.1$_{-0.6}^{+0.9}$    & 1.2$_{-0.1}^{+2.3}$    & 7.5$_{-0.0}^{+5.7}$    & 12.7$_{-2.4}^{+6.0}$       \\
65       & --   & 0.0$_{-0.0}^{+2.7}$    & 0.0$_{-0.0}^{+2.0}$   & 6.0$_{-2.5}^{+2.6}$    & 0.0$_{-0.0}^{+1.0}$    & 21.3$_{-16.6}^{+9.9}$  & 8.6$_{-5.4}^{+5.3}$   & 8.3$_{-3.2}^{+4.2}$    & 0.0$_{-2.7}^{+1.3}$    & 41.8$_{-22.4}^{+15.0}$ & 13.9$_{-10.4}^{+7.9}$      \\
$\cdots$ &      & 54.9$_{-19.1}^{+56.6}$ & 0.1$_{-0.0}^{+7.9}$   & 0.0$_{-0.0}^{+6.3}$    & 8.8$_{-6.6}^{+16.2}$   & 6.2$_{-2.8}^{+9.7}$    & 0.3$_{-0.0}^{+9.6}$   & 0.0$_{-0.0}^{+4.0}$    & 2.7$_{-1.4}^{+3.8}$    & 8.5$_{-2.5}^{+28.2}$   & 18.4$_{-7.7}^{+20.9}$      \\
66       & K5   & 2.1$_{-0.0}^{+8.1}$    & 0.0$_{-0.0}^{+1.6}$   & 0.1$_{-0.0}^{+1.5}$    & 0.8$_{-0.0}^{+2.4}$    & 46.5$_{-15.6}^{+15.4}$ & 0.8$_{-0.0}^{+5.1}$   & 2.9$_{-1.6}^{+5.0}$    & 0.0$_{-0.0}^{+2.3}$    & 38.2$_{-18.7}^{+18.7}$ & 8.7$_{-2.5}^{+8.2}$        \\
$\cdots$ &      & 33.5$_{-16.5}^{+10.5}$ & 3.9$_{-0.0}^{+16.1}$  & 4.2$_{-0.0}^{+14.3}$   & 0.0$_{-0.0}^{+4.9}$    & 26.3$_{-8.9}^{+25.8}$  & 1.7$_{-0.0}^{+11.2}$  & 0.4$_{-0.0}^{+8.1}$    & 3.2$_{-0.0}^{+19.6}$   & 11.0$_{-0.0}^{+21.0}$  & 15.7$_{-8.1}^{+6.8}$       \\
71       & M3   & 6.4$_{-0.0}^{+28.2}$   & 0.0$_{-0.0}^{+1.8}$   & 2.3$_{-0.0}^{+6.8}$    & 0.8$_{-0.0}^{+8.7}$    & 24.2$_{-11.4}^{+24.6}$ & 3.0$_{-0.3}^{+20.1}$  & 4.7$_{-1.6}^{+25.9}$   & 1.9$_{-0.5}^{+12.1}$   & 56.6$_{-10.9}^{+55.6}$ & 0.0$_{-0.0}^{+10.4}$       \\
74       & --   & 0.0$_{-0.0}^{+5.1}$    & 0.0$_{-0.0}^{+1.7}$   & 6.3$_{-2.8}^{+3.7}$    & 13.4$_{-6.3}^{+4.8}$   & 0.0$_{-0.0}^{+6.3}$    & 5.8$_{-3.0}^{+3.4}$   & 16.1$_{-10.0}^{+5.7}$  & 0.0$_{-0.0}^{+4.7}$    & 58.3$_{-31.2}^{+17.9}$ & 0.0$_{-0.0}^{+18.2}$       \\
$\cdots$ &      & 50.7$_{-19.5}^{+33.8}$ & 0.1$_{-0.0}^{+11.1}$  & 0.1$_{-0.0}^{+6.4}$    & 0.0$_{-0.0}^{+29.4}$   & 24.2$_{-9.8}^{+26.5}$  & 1.8$_{-0.0}^{+23.4}$  & 5.4$_{-0.0}^{+31.1}$   & 0.0$_{-0.0}^{+12.8}$   & 9.9$_{-5.9}^{+22.5}$   & 7.8$_{-7.1}^{+23.4}$       \\
75       & --   & 0.0$_{-0.0}^{+72.5}$   & 0.0$_{-0.0}^{+10.2}$  & 4.1$_{-0.0}^{+28.5}$   & 0.1$_{-0.0}^{+20.2}$   & 15.0$_{-0.0}^{+ >15.0}$ & 5.2$_{-0.0}^{+36.0}$  & 0.0$_{-0.0}^{+32.5}$   & 0.0$_{-0.0}^{+51.2}$   & 59.6$_{-33.4}^{+42.6}$ & 16.0$_{-0.0}^{+69.6}$      \\
76       & M1   & 5.3$_{-3.4}^{+9.6}$    & 0.0$_{-0.0}^{+2.8}$   & 2.6$_{-0.0}^{+3.3}$    & 1.9$_{-0.0}^{+3.1}$    & 65.0$_{-29.6}^{+15.7}$ & 0.0$_{-0.0}^{+3.3}$   & 1.3$_{-0.0}^{+5.3}$    & 0.0$_{-0.0}^{+3.0}$    & 23.7$_{-3.9}^{+13.5}$  & 0.1$_{-0.0}^{+2.0}$        \\
$\cdots$ &      & 41.8$_{-23.6}^{+17.5}$ & 3.6$_{-1.9}^{+4.5}$   & 0.0$_{-0.0}^{+3.3}$    & 0.0$_{-0.0}^{+2.8}$    & 29.6$_{-12.1}^{+12.4}$ & 0.0$_{-0.0}^{+3.3}$   & 12.9$_{-5.2}^{+7.2}$   & 0.0$_{-0.0}^{+2.1}$    & 12.2$_{-4.3}^{+5.9}$   & 0.0$_{-1.0}^{+2.1}$        \\
80       & --   & 0.3$_{-0.0}^{+1.2}$    & 0.0$_{-0.0}^{+0.2}$   & 1.2$_{-0.0}^{+0.5}$    & 0.0$_{-0.0}^{+0.4}$    & 7.5$_{-0.0}^{+2.0}$    & 3.2$_{-1.5}^{+1.3}$   & 3.1$_{-0.9}^{+1.0}$    & 0.0$_{-0.0}^{+0.6}$    & 84.7$_{-6.6}^{+8.2}$   & 0.0$_{-0.0}^{+1.1}$        \\
$\cdots$ &      & 6.4$_{-0.0}^{+50.4}$   & 8.3$_{-0.0}^{+29.8}$  & 0.6$_{-0.0}^{+12.4}$   & 7.4$_{-0.0}^{+34.2}$   & 15.4$_{-0.0}^{+50.2}$  & 6.2$_{-0.0}^{+18.2}$  & 0.8$_{-0.0}^{+10.1}$   & 3.1$_{-0.0}^{+9.0}$    & 22.6$_{-0.0}^{+78.6}$  & 29.2$_{-0.0}^{+67.6}$      \\
81       & M5   & 10.7$_{-0.0}^{+2.8}$   & 0.0$_{-0.0}^{+0.8}$   & 0.0$_{-0.0}^{+0.4}$    & 4.3$_{-0.0}^{+1.2}$    & 54.5$_{-3.7}^{+5.4}$   & 0.0$_{-0.0}^{+0.7}$   & 6.1$_{-2.0}^{+1.9}$    & 0.0$_{-0.0}^{+0.5}$    & 15.6$_{-2.7}^{+2.5}$   & 8.8$_{-1.2}^{+1.6}$        \\
$\cdots$ &      & 45.4$_{-5.1}^{+6.2}$   & 0.0$_{-0.0}^{+1.1}$   & 0.0$_{-0.0}^{+0.4}$    & 0.0$_{-0.0}^{+1.5}$    & 31.4$_{-4.4}^{+8.3}$   & 7.0$_{-2.2}^{+2.4}$   & 0.0$_{-0.0}^{+0.3}$    & 4.0$_{-2.0}^{+2.3}$    & 7.7$_{-1.2}^{+1.0}$    & 4.5$_{-0.9}^{+0.7}$        \\
86       & M5.5 & 15.7$_{-1.7}^{+3.1}$   & 0.0$_{-0.0}^{+0.5}$   & 0.8$_{-0.0}^{+0.7}$    & 0.0$_{-0.0}^{+0.2}$    & 36.0$_{-3.0}^{+3.4}$   & 0.0$_{-0.0}^{+0.2}$   & 3.3$_{-1.7}^{+1.1}$    & 0.0$_{-0.0}^{+0.2}$    & 44.1$_{-4.4}^{+4.2}$   & 0.1$_{-0.0}^{+0.7}$        \\
$\cdots$ &      & 26.7$_{-0.0}^{+83.1}$  & 12.2$_{-0.0}^{+28.5}$ & 0.0$_{-0.0}^{+38.4}$   & 1.9$_{-0.0}^{+14.8}$   & 18.2$_{-0.0}^{+55.5}$  & 10.0$_{-0.0}^{+27.7}$ & 0.5$_{-0.0}^{+20.3}$   & 6.0$_{-0.0}^{+29.6}$   & 10.7$_{-0.0}^{+39.6}$  & 13.7$_{-0.0}^{+38.7}$      \\
88       & M0.5 & 0.0$_{-0.0}^{+5.0}$    & 0.0$_{-0.0}^{+1.0}$   & 8.6$_{-0.0}^{+4.4}$    & 0.0$_{-0.0}^{+0.4}$    & 67.1$_{-17.3}^{+20.5}$ & 10.3$_{-5.0}^{+7.1}$  & 0.0$_{-0.0}^{+3.6}$    & 3.1$_{-2.0}^{+2.4}$    & 9.9$_{-1.8}^{+6.3}$    & 0.9$_{-0.6}^{+1.2}$        \\
$\cdots$ &      & 6.0$_{-10.0}^{+30.9}$  & 0.7$_{-9.0}^{+37.8}$  & 33.3$_{-13.7}^{+22.2}$ & 6.0$_{-9.2}^{+31.7}$   & 2.1$_{-6.5}^{+21.3}$   & 0.0$_{-0.0}^{+7.7}$   & 30.6$_{-12.4}^{+29.1}$ & 0.6$_{-3.0}^{+9.9}$    & 17.6$_{-10.1}^{+32.8}$ & 3.1$_{-3.5}^{+11.2}$       \\
89       & K5   & 0.0$_{-0.0}^{+0.7}$    & 0.0$_{-0.0}^{+0.2}$   & 2.6$_{-1.0}^{+1.6}$    & 1.6$_{-0.0}^{+0.7}$    & 0.0$_{-0.0}^{+0.7}$    & 0.0$_{-0.0}^{+0.6}$   & 1.4$_{-0.0}^{+0.9}$    & 0.0$_{-0.0}^{+0.6}$    & 94.4$_{-75.4}^{+18.8}$ & 0.0$_{-0.0}^{+3.7}$        \\
90       & --   & 26.0$_{-12.1}^{+20.9}$ & 0.0$_{-0.1}^{+4.1}$   & 3.0$_{-1.6}^{+4.4}$    & 0.0$_{-0.0}^{+2.7}$    & 49.1$_{-30.7}^{+24.7}$ & 3.7$_{-1.8}^{+9.5}$   & 0.0$_{-0.0}^{+10.1}$   & 2.8$_{-1.9}^{+5.5}$    & 5.4$_{-1.2}^{+16.4}$   & 10.0$_{-1.8}^{+12.9}$      \\
$\cdots$ &      & 31.1$_{-0.0}^{+65.4}$  & 7.5$_{-1.4}^{+13.4}$  & 8.5$_{-0.0}^{+51.3}$   & 0.7$_{-0.0}^{+21.3}$   & 16.8$_{-0.0}^{+54.1}$  & 3.6$_{-0.0}^{+38.8}$  & 9.2$_{-0.0}^{+58.7}$   & 0.8$_{-0.0}^{+24.8}$   & 12.5$_{-0.0}^{+52.2}$  & 9.2$_{-0.0}^{+39.3}$       \\
92       & M0   & 1.1$_{-0.0}^{+9.7}$    & 0.0$_{-0.0}^{+0.9}$   & 4.8$_{-0.0}^{+4.8}$    & 2.3$_{-0.1}^{+5.3}$    & 35.8$_{-8.5}^{+25.4}$  & 5.7$_{-2.1}^{+7.5}$   & 0.0$_{-0.0}^{+3.5}$    & 0.0$_{-0.0}^{+2.4}$    & 44.0$_{-15.9}^{+13.1}$ & 6.3$_{-2.4}^{+14.1}$       \\
$\cdots$ &      & 19.6$_{-7.1}^{+9.7}$   & 0.8$_{-0.0}^{+3.4}$   & 0.2$_{-0.0 }^{+9.0}$   & 3.0$_{-0.0}^{+4.6}$    & 31.2$_{-15.1}^{+9.3}$  & 0.4$_{-0.0}^{+6.5}$   & 6.8$_{-3.9}^{+3.8}$    & 0.0$_{-0.0}^{+10.0}$   & 37.9$_{-12.3}^{+19.4}$ & 0.0$_{-0.0}^{+12.6}$       \\
96       & M1   & 18.2$_{-11.8}^{+8.7}$  & 2.6$_{-2.0}^{+3.8}$   & 2.8$_{-1.6}^{+4.6}$    & 0.0$_{-0.0}^{+5.9}$    & 56.5$_{-35.1}^{+23.3}$ & 3.8$_{-1.8}^{+12.9}$  & 0.0$_{-0.0}^{+15.2}$   & 3.0$_{-0.8}^{+9.2}$    & 7.4$_{-4.3}^{+6.7}$    & 5.8$_{-1.5}^{+10.1}$       \\

\hline
\end{tabular}
\flushleft{$^a$ Amorphous olivine and pyroxene combined.} \\
$^b$ For each object, the first line corresponds to the warm component
abundances, while the second line corresponds to the cold component
abundances, when available.\\
\end{table}

\begin{table}
\centering
\tiny
\caption{Continuation} 
\begin{tabular}{lc|cccc|cccc|cc}
\hline
ID & SpT & Oli/Pyr$^a$ \% & Ens \% & For \% & Sil \% & Oli/Pyr \% & Ens \% & For \% & Sil \% & Oli/Pyr \% & Sil \% \\
   &     & (0.1 $\mu$m) & (0.1 $\mu$m) & (0.1 $\mu$m) & (0.1 $\mu$m) & (1.5 $\mu$m) & (1.5 $\mu$m) & (1.5 $\mu$m) & (1.5 $\mu$m) & (6.0 $\mu$m) & (6.0 $\mu$m) \\
\hline
100      & --   & 0.0$_{-0.0}^{+29.3}$   & 0.0$_{-0.0}^{+12.0}$  & 2.8$_{-0.0}^{+42.3}$   & 0.5$_{-0.0}^{+35.5}$   & 1.2$_{-0.0}^{+88.8}$   & 9.5$_{-0.0}^{+81.9}$  & 0.0$_{-0.0}^{+34.1}$   & 1.1$_{-0.0}^{+106.2}$  & 81.8$_{-54.1}^{+47.3}$ & 3.1$_{-4.1}^{+11.6}$       \\
101      & --   & 24.1$_{-6.9}^{+26.6}$  & 0.0$_{-0.0}^{+4.1}$   & 2.0$_{-1.4}^{+3.2}$    & 1.0$_{-0.0}^{+4.2}$    & 60.2$_{-33.4}^{+22.6}$ & 1.9$_{-0.0}^{+8.8}$   & 2.6$_{-0.5}^{+7.5}$    & 0.0$_{-0.0}^{+5.3}$    & 6.6$_{-0.0}^{+10.1}$   & 1.6$_{-0.0}^{+5.2}$        \\
$\cdots$ &      & 48.0$_{-19.8}^{+7.1}$  & 0.0$_{-0.1}^{+2.7}$   & 0.2$_{-0.0}^{+2.7}$    & 2.5$_{-1.9}^{+3.1}$    & 55.9$_{-11.9}^{+7.1}$  & 0.0$_{-0.0}^{+3.7}$   & 9.5$_{-6.1}^{+4.8}$    & 0.0$_{-0.0}^{+2.7}$    & 13.6$_{-7.2}^{+3.5}$   & 0.1$_{-0.0}^{+3.0}$        \\
103      & --   & 0.0$_{-0.0}^{+0.9}$    & 0.0$_{-0.0}^{+0.3}$   & 2.4$_{-0.0}^{+0.6}$    & 3.2$_{-0.0}^{+0.9}$    & 38.0$_{-5.0}^{+7.0}$   & 2.0$_{-1.2}^{+2.3}$   & 7.8$_{-2.1}^{+2.4}$    & 0.0$_{-0.0}^{+1.8}$    & 27.6$_{-5.5}^{+6.0}$   & 18.9$_{-4.1}^{+4.2}$       \\
$\cdots$ &      & 11.0$_{-9.1}^{+3.7}$   & 5.8$_{-2.7}^{+4.2}$   & 0.0$_{-0.0}^{+3.9}$    & 5.8$_{-1.8}^{+3.3}$    & 35.1$_{-20.5}^{+18.1}$ & 0.0$_{-0.0}^{+6.3}$   & 5.1$_{-0.0}^{+5.6}$    & 4.2$_{-0.0}^{+9.0}$    & 15.7$_{-4.5}^{+8.5}$   & 17.3$_{-17.7}^{+2.3}$      \\
104      & --   & 1.6$_{-1.1}^{+14.7}$   & 0.9$_{-0.2}^{+6.1}$   & 3.7$_{-2.3}^{+5.2}$    & 4.3$_{-1.8}^{+4.1}$    & 70.4$_{-48.8}^{+35.4}$ & 4.7$_{-3.6}^{+12.4}$  & 0.0$_{-0.0}^{+10.2}$   & 0.0$_{-0.0}^{+6.1}$    & 14.2$_{-6.1}^{+17.3}$  & 0.1$_{-0.0}^{+14.8}$       \\
106      & M3   & 0.0$_{-0.0}^{+16.0}$   & 5.1$_{-0.0}^{+23.7}$  & 4.0$_{-0.0}^{+17.1}$   & 5.7$_{-0.0}^{+34.2}$   & 0.1$_{-0.0}^{+24.9}$   & 10.0$_{-5.5}^{+21.9}$ & 0.3$_{-0.0}^{+10.7}$   & 1.1$_{-0.0}^{+15.6}$   & 73.7$_{-61.5}^{+19.9}$ & 0.0$_{-0.0}^{+20.7}$       \\
113      & K7   & 0.0$_{-0.0}^{+1.2}$    & 0.0$_{-0.0}^{+0.7}$   & 3.3$_{-0.5}^{+1.7}$    & 4.4$_{-1.9}^{+1.3}$    & 24.9$_{-10.3}^{+12.9}$ & 10.3$_{-3.4}^{+4.9}$  & 5.8$_{-2.2}^{+5.2}$    & 0.0$_{-0.0}^{+2.0}$    & 42.5$_{-17.9}^{+13.4}$ & 8.8$_{-4.4}^{+7.0}$        \\
$\cdots$ &      & 36.9$_{-14.1}^{+19.9}$ & 2.1$_{-0.0}^{+8.1}$   & 3.6$_{-1.4}^{+4.2}$    & 0.0$_{-0.0}^{+2.6}$    & 36.2$_{-13.5}^{+21.8}$ & 1.7$_{-0.0}^{+9.0}$   & 0.0$_{-0.0}^{+5.8}$    & 0.0$_{-0.0}^{+7.9}$    & 10.9$_{-1.3}^{+11.9}$  & 8.6$_{-4.2}^{+10.8}$       \\
114      & F9   & 0.4$_{-0.0}^{+5.3}$    & 0.0$_{-0.0}^{+0.5}$   & 2.8$_{-1.1}^{+1.2}$    & 0.1$_{-0.0}^{+0.8}$    & 34.7$_{-12.5}^{+17.7}$ & 4.2$_{-2.8}^{+3.2}$   & 1.9$_{-0.6}^{+6.6}$    & 0.0$_{-0.0}^{+2.3}$    & 51.8$_{-17.1}^{+15.2}$ & 17.4$_{-5.2}^{+10.1}$      \\
115      & M0.5 & 3.5$_{-2.2}^{+6.8}$    & 0.4$_{-0.2}^{+1.2}$   & 1.5$_{-0.0}^{+1.5}$    & 0.6$_{-0.0}^{+1.2}$    & 65.7$_{-48.2}^{+17.3}$ & 0.0$_{-0.0}^{+4.3}$   & 0.0$_{-0.0}^{+4.3}$    & 0.0$_{-0.0}^{+3.5}$    & 25.6$_{-14.5}^{+10.3}$ & 2.6$_{-0.9}^{+4.6}$        \\
117      & K2   & 74.6$_{-31.9}^{+32.8}$ & 1.4$_{-0.0}^{+3.6}$   & 5.9$_{-4.1}^{+2.1}$    & 0.0$_{-0.0}^{+1.6}$    & 5.1$_{-0.2}^{+6.4}$    & 10.7$_{-3.2}^{+2.1}$  & 0.0$_{-0.0}^{+0.9}$    & 0.0$_{-0.0}^{+0.9}$    & 2.1$_{-0.0}^{+2.0}$    & 0.2$_{-0.0}^{+1.2}$        \\
$\cdots$ &      & 37.9$_{-15.7}^{+14.9}$ & 6.3$_{-4.7}^{+5.5}$   & 19.0$_{-9.9}^{+9.2}$   & 6.1$_{-4.1}^{+5.3}$    & 23.6$_{-6.8}^{+14.9}$  & 0.8$_{-0.4}^{+2.1}$   & 0.0$_{-0.0}^{+6.8}$    & 2.3$_{-2.4}^{+3.9}$    & 2.4$_{-1.3}^{+2.8}$    & 1.7$_{-0.9}^{+2.2}$        \\
119      & K7   & 0.0$_{-0.0}^{+11.2}$   & 0.7$_{-0.0}^{+23.7}$  & 1.1$_{-0.0}^{+17.4}$   & 1.9$_{-0.0}^{+31.6}$   & 11.1$_{-2.7}^{+56.7}$  & 0.2$_{-0.0}^{+25.1}$  & 0.6$_{-0.0}^{+41.3}$   & 0.3$_{-0.0}^{+33.6}$   & 84.0$_{-26.7}^{+58.6}$ & 0.0$_{-0.0}^{+96.2}$       \\
122      & M0   & 30.7$_{-30.4}^{+3.5}$  & 0.0$_{-0.0}^{+7.8}$   & 1.4$_{-0.0}^{+10.5}$   & 0.0$_{-0.0}^{+7.1}$    & 52.6$_{-35.4}^{+10.1}$ & 6.4$_{-3.1}^{+15.0}$  & 0.0$_{-0.0}^{+10.0}$   & 0.0$_{-0.0}^{+4.4}$    & 1.7$_{-1.0}^{+17.3}$   & 7.1$_{-1.1}^{+14.6}$       \\
$\cdots$ &      & 25.3$_{-12.9}^{+38.5}$ & 0.2$_{-0.6}^{+7.5}$   & 5.5$_{-0.0}^{+19.7}$   & 22.7$_{-14.1}^{+36.9}$ & 27.8$_{-13.3}^{+39.9}$ & 0.9$_{-0.0}^{+11.2}$  & 0.0$_{-0.0}^{+14.3}$   & 0.0$_{-1.7}^{+10.5}$   & 11.8$_{-4.3}^{+13.6}$  & 5.7$_{-0.0}^{+18.0}$       \\
123      & M0   & 42.0$_{-19.7}^{+65.0}$ & 0.7$_{-0.0}^{+3.3}$   & 5.4$_{-3.7}^{+2.7}$    & 0.0$_{-0.0}^{+1.4}$    & 41.2$_{-16.8}^{+11.7}$ & 4.1$_{-0.6}^{+6.4}$   & 0.0$_{-0.0}^{+4.2}$    & 2.7$_{-1.4}^{+3.7}$    & 4.0$_{-0.1}^{+6.8}$    & 0.0$_{-0.0}^{+0.5}$        \\
125      & M0   & 4.9$_{-0.0}^{+13.0}$   & 0.0$_{-0.0}^{+3.3}$   & 2.3$_{-1.6}^{+2.5}$    & 0.0$_{-0.0}^{+1.9}$    & 56.9$_{-23.8}^{+11.9}$ & 0.0$_{-0.0}^{+5.0}$   & 0.0$_{-0.0}^{+7.2}$    & 0.0$_{-0.0}^{+4.4}$    & 30.8$_{-16.9}^{+6.4}$  & 5.0$_{-1.2}^{+11.2}$       \\
$\cdots$ &      & 31.0$_{-20.4}^{+10.3}$ & 11.6$_{-12.8}^{+1.5}$ & 0.0$_{-2.6}^{+0.9}$    & 0.0$_{-0.0}^{+2.5}$    & 29.1$_{-13.7}^{+9.2}$  & 7.0$_{-5.2}^{+4.1}$   & 0.2$_{-0.0}^{+1.8}$    & 0.0$_{-0.0}^{+3.8}$    & 16.0$_{-7.7}^{+6.1}$   & 5.1$_{-3.1}^{+3.2}$        \\
127      & M2   & 1.5$_{-0.0}^{+2.9}$    & 3.4$_{-2.6}^{+1.2}$   & 0.0$_{-0.0}^{+2.0}$    & 0.9$_{-0.4}^{+1.9}$    & 0.0$_{-0.0}^{+2.0}$    & 0.0$_{-0.0}^{+3.1}$   & 1.5$_{-0.0}^{+5.7}$    & 1.3$_{-0.5}^{+4.4}$    & 91.5$_{-21.7}^{+71.4}$ & 0.0$_{-0.0}^{+7.8}$        \\
129      & --   & 0.2$_{-0.0}^{+2.4}$    & 0.0$_{-0.0}^{+0.8}$   & 0.9$_{-0.0}^{+1.7}$    & 0.0$_{-0.0}^{+1.3}$    & 14.7$_{-4.2}^{+7.3}$   & 1.7$_{-0.0}^{+4.2}$   & 0.0$_{-0.0}^{+2.3}$    & 0.0$_{-0.0}^{+0.9}$    & 82.5$_{-55.0}^{+20.0}$ & 0.0$_{-0.0}^{+6.7}$        \\
137      & --   & 40.7$_{-12.0}^{+10.3}$ & 0.8$_{-0.0}^{+6.0}$   & 0.9$_{-0.0}^{+2.9}$    & 0.0$_{-0.0}^{+2.3}$    & 39.0$_{-19.2}^{+10.8}$ & 5.6$_{-4.2}^{+3.4}$   & 5.8$_{-3.6}^{+4.1}$    & 0.0$_{-0.0}^{+3.2}$    & 0.0$_{-0.0}^{+2.3}$    & 7.1$_{-2.1}^{+4.7}$        \\
142      & M4   & 0.7$_{-0.0}^{+6.3}$    & 0.5$_{-0.0}^{+2.4}$   & 2.6$_{-1.1}^{+2.3}$    & 1.3$_{-0.2}^{+2.9}$    & 58.7$_{-33.8}^{+20.7}$ & 4.0$_{-2.6}^{+6.6}$   & 1.9$_{-0.0}^{+7.4}$    & 4.5$_{-2.7}^{+4.9}$    & 25.7$_{-10.1}^{+9.0}$  & 0.1$_{-0.0}^{+6.1}$        \\
144      & --   & 0.0$_{-0.0}^{+5.3}$    & 0.0$_{-0.0}^{+2.1}$   & 6.1$_{-1.3}^{+1.6}$    & 4.4$_{-1.9}^{+2.6}$    & 46.2$_{-23.0}^{+6.0}$  & 11.2$_{-3.3}^{+4.4}$  & 0.0$_{-0.0}^{+3.0}$    & 0.0$_{-0.0}^{+3.0}$    & 32.0$_{-7.4}^{+13.9}$  & 0.0$_{-0.0}^{+2.6}$        \\
$\cdots$ &      & 23.3$_{-9.5}^{+91.2}$  & 22.1$_{-11.6}^{+33.3}$ & 10.9$_{-9.9}^{+26.9}$  & 0.0$_{-0.0}^{+41.9}$  & 7.3$_{-7.8}^{+21.8}$   & 10.5$_{-6.9}^{+20.1}$ & 12.1$_{-6.2}^{+10.9}$  & 0.0$_{-0.0}^{+22.2}$   & 8.3$_{-3.5}^{+19.9}$   & 5.5$_{-3.7}^{+14.3}$       \\
146      & M4   & 6.8$_{-1.1}^{+2.7}$    & 0.0$_{-0.0}^{+0.6}$   & 4.4$_{-1.6}^{+1.2}$    & 2.2$_{-1.1}^{+1.1}$    & 34.1$_{-6.7}^{+4.5}$   & 0.0$_{-0.0}^{+1.1}$   & 4.1$_{-2.3}^{+3.0}$    & 0.0$_{-0.0}^{+1.3}$    & 32.6$_{-7.7}^{+8.3}$   & 15.6$_{-4.7}^{+2.9}$       \\
$\cdots$ &      & 17.2$_{-2.6}^{+6.3}$   & 0.0$_{-0.0}^{+4.2}$   & 0.0$_{-0.0}^{+3.8}$    & 0.0$_{-0.0}^{+4.7}$    & 27.7$_{-8.0}^{+8.1}$   & 0.0$_{-0.0}^{+3.8}$   & 2.2$_{-0.0}^{+1.5}$    & 0.0$_{-0.0}^{+2.6}$    & 52.9$_{-10.5}^{+19.8}$ & 0.0$_{-0.0}^{+8.7}$        \\
147      & --   & 0.0$_{-0.0}^{+4.5}$    & 0.0$_{-0.0}^{+2.9}$   & 3.3$_{-1.3}^{+2.8}$    & 1.3$_{-0.9}^{+1.6}$    & 34.1$_{-10.9}^{+18.9}$ & 9.8$_{-4.6}^{+9.0}$   & 0.0$_{-0.6}^{+3.5}$    & 0.7$_{-0.0}^{+4.8}$    & 50.7$_{-41.6}^{+18.1}$ & 0.0$_{-0.0}^{+15.4}$       \\
148      & K7   & 7.4$_{-0.0}^{+15.4}$   & 2.5$_{-1.3}^{+2.7}$   & 3.5$_{-0.8}^{+3.4}$    & 0.0$_{-0.0}^{+1.0}$    & 70.6$_{-49.9}^{+18.2}$ & 3.7$_{-2.0}^{+8.0}$   & 0.0$_{-0.0}^{+5.3}$    & 2.5$_{-0.0}^{+5.1}$    & 5.1$_{-0.0}^{+8.9}$    & 4.6$_{-1.1}^{+5.1}$        \\
$\cdots$ &      & 3.8$_{-0.0}^{+36.6}$   & 17.8$_{-8.3}^{+16.9}$ & 30.8$_{-11.4}^{+24.5}$ & 1.9$_{-0.0}^{+10.7}$   & 2.0$_{-0.0}^{+13.4}$   & 10.9$_{-7.5}^{+16.2}$ & 5.0$_{-3.7}^{+17.7}$   & 2.9$_{-0.0}^{+21.5}$   & 24.7$_{-9.6}^{+18.4}$  & 0.1$_{-0.0}^{+4.7}$        \\
149      & M0   & 37.8$_{-13.3}^{+22.2}$ & 0.7$_{-0.0}^{+2.9}$   & 1.4$_{-0.9}^{+2.8}$    & 0.0$_{-0.0}^{+1.6}$    & 24.1$_{-10.7}^{+10.8}$ & 3.9$_{-1.6}^{+5.5}$   & 0.0$_{-0.0}^{+2.3}$    & 1.3$_{-0.0}^{+2.6}$    & 30.1$_{-8.9}^{+11.2}$  & 0.4$_{-0.0}^{+8.8}$        \\
$\cdots$ &      & 41.7$_{-28.1}^{+24.7}$ & 0.0$_{-0.0}^{+5.5}$   & 4.1$_{-1.5}^{+10.3}$   & 4.4$_{-3.6}^{+8.6}$    & 28.3$_{-21.1}^{+14.5}$ & 3.7$_{-0.5}^{+13.2}$  & 1.1$_{-0.1}^{+12.7}$   & 0.0$_{-0.0}^{+8.3}$    & 8.8$_{-1.2}^{+6.5}$    & 7.8$_{-1.9}^{+8.4}$        \\
\hline
\multicolumn{12}{c}{Taurus} \\
04108+2910   & M0   &  0.1$^{+0.4}_{-0.0}$    & 0.5$^{+1.4}_{-0.0}$    & 2.9$^{+1.2}_{-1.0}$    & 0.0$^{+0.8}_{-0.0}$  & 10.4$^{+4.1}_{-6.1}$   & 11.2$^{+3.1}_{-4.7}$   & 3.6$^{+3.1}_{-1.6}$    & 0.0$^{+0.9}_{-0.0}$  & 71.3$^{+16.2}_{-29.0}$ & 0.0$^{+8.4}_{-0.0}$     \\
$\cdots$     &      & 13.2$^{+9.3}_{-5.6}$   & 0.1$^{+6.5}_{-0.0}$    & 4.0$^{+10.8}_{-0.0}$   & 0.0$^{+9.9}_{-0.0}$  & 58.5$^{+22.3}_{-17.8}$ & 0.0$^{+8.0}_{-0.0}$    & 0.0$^{+7.8}_{-0.0}$    & 0.0$^{+45.4}_{-0.0}$ & 12.9$^{+15.9}_{-5.3}$  & 11.4$^{+7.1}_{-6.4}$    \\
04200+2759   & --   &  4.7$^{+1.5}_{-2.4}$    & 0.1$^{+2.2}_{-0.0}$    & 0.4$^{+1.2}_{-0.0}$    & 0.0$^{+0.9}_{-0.0}$  & 48.4$^{+14.7}_{-13.1}$ & 7.9$^{+6.2}_{-4.1}$    & 2.8$^{+5.3}_{-0.0}$    & 0.0$^{+2.5}_{-0.0}$  & 19.6$^{9.4}_{-5.6}$    & 16.0$^{+7.1}_{-11.1}$    \\
$\cdots$     &      & 14.3$^{+11.5}_{-7.3}$  & 1.2$^{+4.5}_{-7.9}$    & 4.2$^{+5.1}_{-3.6}$    & 3.8$^{+8.2}_{-7.2}$   & 58.6$^{+32.2}_{-26.4}$ & 0.0$^{+9.8}_{-0.0}$    & 0.0$^{+18.6}_{-0.0}$   & 7.2$^{+31.8}_{-20.3}$ & 10.3$^{+7.3}_{-5.6}$   & 0.4$^{+2.6}_{-1.4}$    \\
04216+2603   & M1   & 43.9$^{+15.3}_{-22.8}$ & 1.8$^{+5.4}_{-0.7}$    & 2.1$^{+3.0}_{-1.0}$    & 0.0$^{+2.2}_{-0.0}$  & 18.9$^{+17.6}_{-12.3}$ & 6.4$^{+1.7}_{-5.1}$    & 0.0$^{+2.5}_{-0.2}$    & 0.0$^{+2.8}_{-0.0}$  & 27.1$^{+9.7}_{-9.5}$   & 0.0$^{+9.3}_{-0.0}$      \\
$\cdots$     &      & 18.6$^{+27.5}_{-9.0}$  & 0.1$^{+16.8}_{-0.0}$   & 13.2$^{+31.5}_{-10.7}$ & 1.8$^{+40.3}_{-8.2}$  & 4.4$^{+16.5}_{-5.1}$   & 2.1$^{+18.1}_{-3.1}$   & 1.5$^{+20.5}_{-2.9}$   & 10.5$^{+17.2}_{-6.0}$ & 44.5$^{+31.4}_{-17.3}$ & 3.2$^{+13.5}_{-4.3}$   \\
04303+2240   & --   &  2.4$^{+9.1}_{-0.0}$    & 0.0$^{+1.7}_{-0.0}$    & 2.7$^{+3.6}_{-0.0}$    & 0.0$^{+0.7}_{-0.0}$  & 74.2$^{+25.2}_{-17.9}$ & 3.5$^{+8.5}_{-0.0}$    & 9.6$^{+10.7}_{-6.3}$   & 2.0$^{+4.0}_{-0.4}$  & 0.0$^{+1.0}_{-0.0}$    & 5.5$^{+9.8}_{-0.0}$      \\
04370+2559   & --   &  3.1$^{+7.3}_{-0.0}$    & 1.2$^{+2.3}_{-0.0}$    & 0.7$^{+1.5}_{-0.0}$    & 0.0$^{+0.4}_{-0.0}$  & 80.2$^{+16.5}_{-33.6}$ & 0.3$^{+7.0}_{-0.0}$    & 1.2$^{+5.7}_{-0.0}$    & 0.0$^{+2.1}_{-0.0}$  & 4.9$^{+5.3}_{-0.0}$    & 8.3$^{+2.8}_{-2.3}$      \\
04385+2550   & M0   & 16.2$^{+2.3}_{-7.9}$   & 0.8$^{+1.3}_{-0.5}$    & 1.0$^{+1.9}_{-0.0}$    & 0.0$^{+0.7}_{-0.0}$  & 55.1$^{+12.3}_{-12.6}$ & 4.8$^{+4.9}_{-3.0}$    & 3.7$^{+3.8}_{-2.4}$    & 0.0$^{+2.5}_{-0.0}$  & 6.4$^{+1.9}_{-1.7}$    & 11.9$^{+2.7}_{-6.2}$     \\
$\cdots$     &      & 13.2$^{+11.7}_{-0.0}$  & 0.0$^{+2.2}_{-0.0}$    & 1.6$^{+3.3}_{-0.0}$    & 5.7$^{+11.4}_{-0.0}$  & 47.7$^{+34.5}_{-12.4}$ & 0.0$^{+16.4}_{-0.0}$   & 0.0$^{+9.1}_{-0.0}$    & 19.9$^{+39.6}_{-0.0}$ & 8.1$^{+5.0}_{-1.6}$    & 3.9$^{+6.0}_{-0.0}$    \\
AATau        & K7   & 12.5$^{+10.1}_{-0.6}$  & 0.4$^{+1.0}_{-0.0}$    & 1.4$^{+1.4}_{-0.7}$    & 0.0$^{+0.7}_{-0.0}$  & 30.6$^{+7.5}_{-8.4}$   & 3.4$^{+2.6}_{-2.6}$    & 2.6$^{+3.3}_{-1.6}$    & 2.1$^{+2.6}_{-0.8}$  & 45.7$^{+13.6}_{-20.6}$ & 1.3$^{+5.3}_{-0.6}$      \\
$\cdots$     &      &  2.6$^{+19.7}_{-0.0}$   & 19.2$^{+15.3}_{-0.0}$  & 0.0$^{+24.2}_{-0.0}$   & 18.3$^{+22.4}_{-0.0}$ & 9.3$^{+20.9}_{-0.0}$   & 8.3$^{+16.8}_{-0.0}$   & 4.5$^{+6.0}_{-0.0}$    & 7.0$^{+13.7}_{-0.0}$  & 29.6$^{+37.2}_{-0.0}$  & 1.2$^{+24.4}_{-0.0}$   \\
BPTau	     & K7   & 29.9$^{+5.4}_{-7.9}$   & 1.1$^{+0.9}_{-0.6}$    & 1.0$^{+1.3}_{-0.0}$    & 0.0$^{+0.3}_{-0.0}$  & 25.4$^{+5.6}_{-8.4}$   & 0.8$^{+1.1}_{-0.0}$    & 0.0$^{+0.8}_{-0.0}$    & 0.0$^{+0.5}_{-0.0}$  & 31.6$^{+6.8}_{-10.3}$  & 10.3$^{+2.9}_{-4.1}$     \\
CITau	     & K7   &  2.3$^{+5.6}_{-0.0}$    & 0.0$^{+0.3}_{-0.0}$    & 2.1$^{+1.0}_{-0.0}$    & 0.0$^{+0.9}_{-0.0}$  & 51.8$^{+10.5}_{-12.3}$ & 4.6$^{+6.6}_{-1.0}$    & 2.8$^{+3.7}_{-1.3}$    & 0.1$^{+1.5}_{-0.0}$  & 17.8$^{+13.8}_{-7.8}$  & 18.7$^{+5.1}_{-3.9}$     \\
$\cdots$     &      & 14.8$^{+12.4}_{-7.1}$  & 0.3$^{+1.5}_{-0.0}$    & 3.5$^{+10.3}_{-0.0}$   & 7.5$^{+9.9}_{-0.0}$  & 14.5$^{+12.3}_{-4.3}$  & 1.9$^{+19.1}_{-0.0}$   & 0.0$^{+3.4}_{-0.0}$    & 9.7$^{+7.2}_{-5.8}$  & 27.4$^{+11.9}_{-11.2}$ & 20.3$^{+36.9}_{-6.0}$    \\
CWTau	     & K3   &  8.2$^{+1.9}_{-2.3}$    & 1.0$^{+0.7}_{-0.0}$    & 1.0$^{+1.5}_{-0.0}$    & 0.2$^{+2.4}_{-0.0}$  & 23.5$^{+10.5}_{-5.3}$  & 1.2$^{+3.5}_{-0.0}$    & 0.0$^{+1.1}_{-0.0}$    & 0.0$^{+0.8}_{-0.0}$  & 46.4$^{+9.9}_{-19.3}$  & 18.5$^{+7.4}_{-4.7}$     \\
CoKuTau3     & M1   &  3.3$^{+6.8}_{-0.5}$    & 0.0$^{+1.0}_{-0.0}$    & 2.7$^{+1.6}_{-1.4}$    & 0.0$^{+0.3}_{-0.0}$  & 63.5$^{+15.4}_{-24.8}$ & 3.7$^{+4.5}_{-2.0}$    & 0.4$^{+3.2}_{-0.0}$    & 0.0$^{+2.0}_{-0.0}$  & 26.0$^{+8.3}_{-4.4}$   & 0.3$^{+4.4}_{-0.0}$      \\
$\cdots$     &      & 15.2$^{+9.6}_{-6.0}$   & 5.3$^{+10.2}_{-0.0}$   & 3.8$^{+1.9}_{-2.7}$    & 0.2$^{+4.6}_{-0.0}$  & 24.1$^{+10.5}_{-8.6}$  & 1.8$^{+3.0}_{-0.0}$    & 14.3$^{+25.3}_{-4.3}$  & 4.4$^{+7.3}_{-2.8}$  & 22.5$^{+15.4}_{-10.1}$ & 9.0$^{+4.9}_{-4.9}$      \\
CoKuTau4     & M1.5 & 24.5$^{+3.5}_{-17.1}$  & 0.0$^{+2.7}_{-0.0}$    & 1.0$^{+2.2}_{-0.0}$    & 0.0$^{+1.0}_{-0.0}$  & 44.8$^{+10.8}_{-12.1}$ & 3.7$^{+3.5}_{-1.4}$    & 3.9$^{+3.9}_{-2.8}$    & 2.6$^{+4.5}_{-2.3}$  & 6.0$^{+6.9}_{-0.6}$    & 3.6$^{+4.2}_{-7.1}$      \\
$\cdots$     &      & 13.4$^{+8.4}_{-5.4}$   & 0.3$^{+1.3}_{-0.6}$    & 0.0$^{+1.4}_{-0.7}$    & 13.5$^{+9.1}_{-6.3}$ & 5.8$^{+4.0}_{-2.6}$    & 0.0$^{+0.7}_{-0.4}$    & 0.1$^{+0.7}_{-0.2}$    & 9.5$^{+6.4}_{-4.3}$  & 53.9$^{+26.6}_{-17.1}$ & 3.5$^{+5.3}_{-3.2}$      \\
DDTau        & M3   & 10.6$^{+4.7}_{-2.7}$   & 2.6$^{+3.8}_{-0.0}$    & 2.5$^{+2.6}_{-0.0}$    & 0.0$^{+1.4}_{-0.0}$  & 50.2$^{+19.1}_{-24.5}$ & 5.2$^{+5.6}_{-3.0}$    & 2.1$^{+5.1}_{-1.6}$    & 1.1$^{+4.2}_{-0.6}$  & 1.7$^{+11.5}_{-0.3}$   & 22.9$^{+9.0}_{-9.0}$     \\
$\cdots$     &      & 38.7$^{+26.0}_{-23.6}$ & 9.0$^{+8.8}_{-6.1}$    & 18.5$^{+11.9}_{-8.6}$ & 0.0$^{+5.5}_{-0.0}$  & 16.1$^{+9.4}_{-7.1}$   & 0.1$^{+1.1}_{-0.2}$    & 0.0$^{+2.2}_{-1.2}$    & 0.2$^{+3.9}_{-0.9}$  & 14.5$^{+4.7}_{-5.6}$   & 0.0$^{+1.1}_{-0.0}$      \\
DETau        & M1   & 11.0$^{+6.9}_{-2.4}$   & 0.4$^{+1.7}_{-0.0}$    & 3.3$^{+2.5}_{-0.0}$    & 0.0$^{+0.6}_{-0.0}$  & 37.3$^{+6.9}_{-19.5}$  & 7.2$^{+3.8}_{-2.2}$    & 1.1$^{+3.2}_{-1.1}$    & 0.0$^{+1.4}_{-0.0}$  & 31.3$^{+6.3}_{-6.1}$   & 8.5$^{+5.4}_{-3.3}$      \\
$\cdots$     &      & 51.2$^{+22.1}_{-24.7}$ & 0.3$^{+6.2}_{-0.0}$    & 4.4$^{+6.9}_{-0.0}$    & 0.7$^{+10.2}_{-0.0}$ & 15.1$^{+11.0}_{-4.2}$  & 0.0$^{+5.3}_{-0.0}$    & 0.0$^{+2.5}_{-0.0}$    & 9.3$^{+15.2}_{-0.0}$ & 18.9$^{+14.7}_{-5.2}$  & 0.0$^{+4.6}_{-0.0}$      \\
DFTau	     & M0.5 &  0.6$^{+8.0}_{-0.0}$    & 0.2$^{+2.3}_{-0.0}$    & 1.8$^{+2.1}_{-0.0}$    & 0.3$^{+3.1}_{-0.0}$  & 19.2$^{+3.8}_{-7.8}$   & 3.7$^{+3.9}_{-0.0}$    & 0.0$^{+4.9}_{-0.0}$    & 2.8$^{+6.5}_{-0.0}$  & 71.3$^{+18.8}_{-17.4}$ & 0.0$^{+7.3}_{-0.0}$      \\
DHTau	     & M2   &  2.3$^{+2.1}_{-0.3}$    & 1.8$^{+1.8}_{-0.8}$    & 5.4$^{+2.4}_{-1.2}$    & 1.5$^{+1.6}_{-0.9}$  & 29.6$^{+9.8}_{-10.3}$  & 11.9$^{+5.6}_{-3.1}$   & 0.0$^{+2.2}_{-0.4}$    & 1.1$^{+2.2}_{-1.0}$  & 44.6$^{+17.1}_{-22.4}$ & 0.0$^{+9.9}_{-0.0}$      \\
$\cdots$     &      & 32.0$^{+16.2}_{-12.3}$ & 2.5$^{+6.0}_{-1.7}$    & 3.4$^{+8.5}_{-3.1}$    & 0.0$^{+5.6}_{-1.3}$  & 17.3$^{+20.6}_{-11.8}$ & 5.7$^{+17.2}_{-5.9}$   & 4.3$^{+8.0}_{-3.2}$    & 0.0$^{+22.9}_{-7.3}$ & 22.4$^{+9.4}_{-8.0}$   & 12.4$^{+10.2}_{-7.4}$    \\
DKTau        & M0   & 13.0$^{+3.8}_{-3.1}$   & 0.6$^{+0.9}_{-0.0}$    & 1.9$^{+1.3}_{-1.2}$    & 3.3$^{+1.3}_{-0.9}$  & 54.6$^{+10.1}_{-12.9}$ & 3.1$^{+2.7}_{-1.9}$    & 0.0$^{+1.2}_{-0.0}$    & 1.8$^{+1.8}_{-0.0}$  & 17.2$^{+3.1}_{-3.7}$   & 4.2$^{+1.4}_{-1.1}$      \\
DLTau	     & K7   &  0.1$^{+5.6}_{-1.5}$    & 0.0$^{+1.1}_{-0.0}$    & 3.0$^{+2.3}_{-1.2}$    & 5.0$^{+3.5}_{-2.2}$  & 38.0$^{+17.1}_{-25.0}$ & 10.1$^{+10.2}_{-5.6}$  & 0.0$^{+6.0}_{-0.0}$    & 8.8$^{+7.6}_{-4.7}$  & 24.4$^{+12.3}_{-8.5}$  & 10.7$^{+7.4}_{-4.7}$     \\
DMTau	     & M1   &  4.1$^{+13.6}_{-0.0}$   & 0.0$^{+1.8}_{-0.0}$    & 4.7$^{+2.3}_{-1.0}$    & 0.0$^{+1.2}_{-0.0}$  & 18.0$^{+8.5}_{-13.8}$  & 16.6$^{+6.6}_{-7.4}$   & 2.8$^{+5.4}_{-1.4}$    & 0.0$^{+2.1}_{-0.0}$  & 36.8$^{+14.4}_{-10.8}$ & 17.0$^{+8.3}_{-12.1}$    \\
$\cdots$     &      & 42.6$^{+28.1}_{-17.1}$ & 0.7$^{+8.9}_{-0.0}$    & 1.2$^{+3.6}_{-0.0}$    & 5.3$^{+12.2}_{-4.8}$ & 2.0$^{+2.6}_{-1.0}$    & 0.8$^{+2.8}_{-0.0}$    & 2.1$^{+5.9}_{-0.0}$    & 8.8$^{+9.7}_{-4.7}$  & 29.7$^{+21.5}_{-10.3}$ & 6.8$^{+4.2}_{-5.0}$      \\
DNTau	     & M0   &  3.4$^{+5.1}_{-0.0}$    & 0.0$^{+0.6}_{-0.0}$    & 2.2$^{+1.4}_{-0.6}$    & 1.4$^{+1.1}_{-0.9}$  & 27.6$^{+11.4}_{-3.5}$  & 3.1$^{+2.7}_{-2.0}$    & 1.4$^{+1.9}_{-0.9}$    & 0.0$^{+2.1}_{-0.0}$  & 43.8$^{+23.9}_{-26.5}$ & 17.0$^{+13.5}_{-4.4}$    \\
$\cdots$     &      & 33.7$^{+39.1}_{-0.0}$  & 0.2$^{+9.1}_{-0.0}$    & 0.0$^{+18.1}_{-0.0}$   & 0.0$^{+7.6}_{-0.0}$   & 46.4$^{+53.9}_{-0.0}$  & 0.0$^{+38.7}_{-0.0}$   & 3.6$^{+16.8}_{-0.0}$   & 0.0$^{+ >1.0}_{-0.0}$ & 3.2$^{+11.2}_{-0.0}$   & 12.9$^{+32.7}_{-0.0}$  \\

\hline
\end{tabular}
\flushleft{$^a$ Amorphous olivine and pyroxene combined.} \\
$^b$ For each object, the first line corresponds to the warm component
abundances, while the second line corresponds to the cold component
abundances, when available.\\
\end{table}

\begin{table}
\centering
\tiny
\caption{Continuation} 
\begin{tabular}{lc|cccc|cccc|cc}
\hline
ID & SpT & Oli/Pyr$^a$ \% & Ens \% & For \% & Sil \% & Oli/Pyr \% & Ens \% & For \% & Sil \% & Oli/Pyr \% & Sil \% \\
   &     & (0.1 $\mu$m) & (0.1 $\mu$m) & (0.1 $\mu$m) & (0.1 $\mu$m) & (1.5 $\mu$m) & (1.5 $\mu$m) & (1.5 $\mu$m) & (1.5 $\mu$m) & (6.0 $\mu$m) & (6.0 $\mu$m) \\
\hline
DOTau	     & M0   &  1.2$^{+8.4}_{-0.0}$    & 0.0$^{+3.8}_{-0.0}$    & 1.6$^{+2.7}_{-0.0}$    & 0.3$^{+4.0}_{-0.0}$  & 52.2$^{+28.2}_{-9.1}$  & 4.5$^{+5.7}_{-0.0}$    & 5.3$^{+6.3}_{-0.0}$    & 1.9$^{+6.2}_{-0.0}$  & 19.9$^{+14.9}_{-7.5}$  & 13.1$^{+5.0}_{-4.6}$     \\
DPTau	     & M0.5 & 35.1$^{+13.7}_{-15.8}$ & 0.0$^{+1.5}_{-0.0}$    & 1.1$^{+1.9}_{-0.0}$    & 0.0$^{+1.6}_{-0.0}$  & 50.7$^{+27.5}_{-15.8}$ & 0.3$^{+2.8}_{-0.0}$    & 1.5$^{+4.0}_{-0.0}$    & 0.0$^{+1.8}_{-0.0}$  & 0.2$^{+1.6}_{-0.0}$    & 11.0$^{+5.6}_{-0.0}$     \\
$\cdots$     &      & 20.1$^{+8.9}_{-8.0}$   & 9.8$^{+4.6}_{-5.1}$    & 3.0$^{+3.7}_{-2.1}$    & 7.7$^{+3.8}_{-3.8}$  & 44.4$^{+18.0}_{-23.3}$ & 0.0$^{+5.4}_{-1.6}$    & 7.8$^{+5.1}_{-3.8}$    & 0.0$^{+2.7}_{-0.3}$  & 4.0$^{+1.2}_{-1.0}$    & 3.1$^{+2.3}_{-0.0}$      \\
DQTau	     & M0   &  1.8$^{+1.2}_{-1.3}$    & 5.5$^{+1.7}_{-1.4}$    & 2.0$^{+1.1}_{-0.7}$    & 0.0$^{+0.6}_{-0.0}$  & 20.6$^{+6.8}_{-8.1}$   & 6.0$^{+3.1}_{-2.0}$    & 0.0$^{+1.4}_{-0.0}$    & 1.7$^{+2.8}_{-1.0}$  & 62.3$^{+22.9}_{-34.4}$ & 0.0$^{+5.8}_{-0.0}$      \\
$\cdots$     &      & 14.4$^{+12.0}_{-5.6}$   & 0.1$^{+8.7}_{-0.0}$    & 0.0$^{+8.0}_{-0.0}$    & 0.0$^{+3.3}_{-0.0}$  & 30.8$^{+36.9}_{-14.5}$ & 0.0$^{+12.6}_{-0.0}$   & 4.6$^{+8.4}_{-3.1}$    & 0.0$^{+11.7}_{-0.2}$ & 21.1$^{+36.2}_{-10.5}$ & 29.1$^{+21.2}_{-11.9}$   \\
DRTau        & K7   & 25.0$^{+5.8}_{-5.3}$   & 0.0$^{+0.7}_{-0.0}$    & 1.0$^{+1.5}_{-0.0}$    & 0.0$^{+0.6}_{-0.0}$  & 31.6$^{+9.3}_{-8.7}$   & 0.6$^{+1.5}_{-0.0}$    & 1.6$^{+2.1}_{-1.1}$    & 0.0$^{+1.0}_{-0.0}$  & 20.4$^{+9.9}_{-9.6}$   & 19.8$^{+6.4}_{-4.7}$     \\
DSTau        & K5   & 35.9$^{+7.1}_{-6.3}$   & 0.8$^{+2.9}_{-0.7}$    & 2.8$^{+2.3}_{-1.7}$    & 0.2$^{+1.2}_{-0.0}$  & 17.4$^{+13.7}_{-4.1}$  & 6.2$^{+3.4}_{-3.1}$    & 0.2$^{+1.4}_{-0.0}$    & 0.0$^{+1.1}_{-0.0}$  & 36.4$^{+11.9}_{-10.5}$ & 0.0$^{+4.5}_{-0.0}$      \\
$\cdots$     &      & 22.0$^{+14.7}_{-10.3}$ & 0.2$^{+3.5}_{-1.8}$    & 2.7$^{+3.3}_{-1.8}$    & 17.2$^{+14.0}_{-9.7}$ & 13.4$^{+10.9}_{-6.7}$  & 20.1$^{+17.0}_{-10.8}$ & 0.1$^{+3.6}_{-1.4}$    & 7.0$^{+7.3}_{-4.8}$   & 11.0$^{+7.6}_{-4.9}$   & 6.3$^{+6.0}_{-4.1}$    \\
F04147+2822  & M4   & 10.2$^{+6.3}_{-0.0}$   & 0.0$^{+4.3}_{-0.0}$    & 2.9$^{+2.5}_{-0.0}$    & 0.0$^{+1.4}_{-0.0}$  & 71.2$^{+11.8}_{-25.9}$ & 0.0$^{+4.4}_{-0.0}$    & 0.0$^{+4.0}_{-0.0}$    & 0.1$^{+2.8}_{-0.0}$  & 3.6$^{+3.4}_{-0.1}$    & 12.0$^{+5.3}_{-3.4}$     \\
F04192+2647  & --   &  3.4$^{+12.4}_{-0.0}$   & 0.1$^{+0.8}_{-0.0}$    & 3.6$^{+1.7}_{-0.8}$    & 0.0$^{+0.7}_{-0.0}$  & 49.0$^{+10.0}_{-11.6}$ & 8.7$^{+3.6}_{-4.0}$    & 0.9$^{+3.3}_{-0.5}$    & 0.7$^{+4.2}_{-0.1}$  & 21.5$^{+11.8}_{-9.9}$  & 12.1$^{+4.2}_{-4.0}$     \\
$\cdots$     &      & 19.7$^{+16.8}_{-7.4}$  & 0.0$^{+9.0}_{-1.6}$    & 14.2$^{+15.3}_{-7.1}$  & 3.7$^{+9.4}_{-4.5}$  & 16.1$^{+12.1}_{-6.0}$  & 5.3$^{+9.9}_{-3.2}$    & 11.7$^{+16.0}_{-7.1}$  & 0.0$^{+5.8}_{-0.0}$  & 19.4$^{+12.2}_{-7.1}$  & 9.7$^{+8.5}_{-4.8}$      \\
F04262+2654  & --   &  0.0$^{+1.0}_{-0.0}$    & 2.1$^{+2.0}_{-0.9}$    & 0.0$^{+1.2}_{-0.0}$    & 3.9$^{+2.8}_{-2.4}$  & 53.7$^{+59.3}_{-16.5}$ & 14.9$^{+14.3}_{-5.7}$  & 0.0$^{+8.6}_{-0.0}$    & 3.7$^{+9.6}_{-3.1}$  & 19.1$^{+12.1}_{-8.9}$  & 2.6$^{+7.6}_{-2.2}$      \\
F04297+2246  & --   &  5.7$^{+3.7}_{-1.3}$    & 0.0$^{+1.5}_{-0.0}$    & 2.6$^{+2.0}_{-1.5}$    & 0.3$^{+1.6}_{-0.0}$  & 34.0$^{+7.8}_{-16.4}$  & 9.5$^{+4.4}_{-2.9}$    & 2.8$^{+4.4}_{-1.9}$    & 0.0$^{+3.2}_{-0.0}$  & 44.1$^{+9.8}_{-26.1}$  & 0.9$^{+12.5}_{-1.0}$     \\
$\cdots$     &      & 24.7$^{+61.7}_{-6.4}$  & 3.5$^{+15.1}_{-0.0}$   & 0.1$^{+32.3}_{-0.0}$   & 0.0$^{+17.6}_{-0.0}$ & 22.8$^{+66.2}_{-0.0}$  & 0.0$^{+26.4}_{-0.0}$   & 18.7$^{+18.5}_{-15.3}$ & 5.4$^{+35.6}_{-0.0}$ & 21.3$^{+29.8}_{-5.7}$  & 3.5$^{+23.3}_{-0.0}$     \\
F04297+2246A & --   &  6.0$^{+4.4}_{-1.5}$    & 3.6$^{+1.8}_{-1.8}$    & 3.1$^{+1.0}_{-0.7}$    & 0.0$^{+0.9}_{-0.0}$  & 45.9$^{+9.6}_{-10.3}$  & 9.0$^{+2.9}_{-3.1}$    & 3.4$^{+3.0}_{-2.4}$    & 2.9$^{+2.3}_{-1.5}$  & 23.9$^{+4.1}_{-5.3}$   & 2.1$^{+8.0}_{-0.0}$      \\
$\cdots$     &      & 31.3$^{+15.2}_{-10.4}$ & 1.4$^{+2.8}_{-0.0}$    & 2.1$^{+2.1}_{-0.0}$    & 0.0$^{+8.2}_{-0.0}$  & 20.1$^{+11.1}_{-9.1}$  & 0.2$^{+2.4}_{-0.0}$    & 4.3$^{+5.4}_{-0.0}$    & 0.0$^{+1.4}_{-0.0}$  & 11.2$^{+6.3}_{-5.2}$   & 29.2$^{+15.0}_{-15.7}$   \\
F04570+2520  & --   &  2.9$^{+11.6}_{-0.0}$   & 0.1$^{+1.7}_{-0.0}$    & 0.1$^{+3.9}_{-0.0}$    & 0.0$^{+2.9}_{-0.0}$  & 64.1$^{+24.3}_{-22.8}$ & 5.8$^{+6.2}_{-5.2}$    & 6.6$^{+6.4}_{-3.9}$    & 0.0$^{+4.7}_{-0.0}$  & 20.5$^{+11.0}_{-4.3}$  & 0.0$^{+3.2}_{-0.3}$      \\
$\cdots$     &      &  0.1$^{+4.1}_{-8.8}$    & 0.0$^{+1.8}_{-1.9}$    & 0.0$^{+9.0}_{-0.0}$    & 40.0$^{+8.3}_{-30.3}$ & 0.0$^{+3.6}_{-0.9}$    & 0.0$^{+7.3}_{-0.0}$    & 13.7$^{+9.1}_{-13.9}$  & 15.7$^{+7.4}_{-12.3}$ & 21.5$^{+9.9}_{-17.6}$  & 9.0$^{+2.6}_{-7.4}$    \\
FMTau        & M0   & 56.0$^{+14.4}_{-10.3}$ & 0.0$^{+2.4}_{-0.0}$    & 2.1$^{+2.6}_{-1.5}$    & 0.0$^{+1.7}_{-0.0}$  & 12.4$^{+2.5}_{-9.9}$   & 8.2$^{+3.2}_{-3.6}$    & 3.3$^{+6.0}_{-1.4}$    & 0.0$^{+1.5}_{-0.0}$  & 13.0$^{+5.8}_{-5.4}$   & 5.0$^{+7.4}_{-0.7}$      \\
$\cdots$     &      & 25.0$^{+46.8}_{-0.0}$  & 2.4$^{+23.7}_{-0.0}$   & 3.5$^{+4.5}_{-0.0}$    & 1.2$^{+17.8}_{-0.0}$ & 6.4$^{+10.5}_{-0.0}$   & 1.1$^{+3.7}_{-0.0}$    & 1.3$^{+17.3}_{-0.0}$   & 5.4$^{+14.0}_{-0.0}$ & 28.3$^{+17.3}_{-9.4}$  & 25.4$^{+45.6}_{-7.9}$    \\
FNTau        & M5   & 11.3$^{+3.5}_{-2.0}$   & 4.5$^{+2.7}_{-1.4}$    & 2.7$^{+2.0}_{-1.0}$    & 0.1$^{+2.0}_{-0.2}$  & 47.1$^{+7.4}_{-14.4}$  & 10.9$^{+5.0}_{-3.7}$   & 0.1$^{+3.9}_{-0.0}$    & 3.2$^{+3.2}_{-2.3}$  & 20.1$^{+2.7}_{-7.2}$   & 0.0$^{+14.5}_{-0.0}$     \\
$\cdots$     &      & 10.6$^{+52.1}_{-0.0}$  & 9.3$^{+29.6}_{-0.0}$   & 12.6$^{+37.1}_{-0.0}$  & 0.0$^{+22.9}_{-0.0}$ & 19.0$^{+68.3}_{-0.0}$  & 1.3$^{+16.2}_{-0.0}$   & 2.9$^{+14.3}_{-0.0}$   & 1.8$^{+36.4}_{-0.0}$ & 18.8$^{+56.1}_{-0.0}$  & 23.6$^{+102.6}_{-0.0}$   \\
FOTau        & M2   & 12.4$^{+6.8}_{-6.2}$   & 0.0$^{+4.2}_{-0.0}$    & 2.5$^{+6.5}_{-0.0}$    & 0.2$^{+3.3}_{-0.0}$  & 20.6$^{+10.7}_{-12.0}$ & 6.6$^{+10.5}_{-2.3}$   & 0.0$^{+5.4}_{-0.0}$    & 5.6$^{+13.5}_{-0.0}$ & 52.1$^{+23.4}_{-18.6}$ & 0.0$^{+19.4}_{-0.0}$     \\
FPTau        & M4   & 25.8$^{+9.3}_{-5.8}$   & 3.7$^{+2.6}_{-2.4}$    & 0.9$^{+1.6}_{-0.4}$    & 0.0$^{+2.4}_{-0.0}$  & 5.5$^{+11.8}_{-3.6}$   & 0.0$^{+3.2}_{-0.2}$    & 0.0$^{+1.7}_{-0.0}$    & 0.0$^{+1.1}_{-0.0}$  & 64.1$^{+25.1}_{-20.7}$ & 0.0$^{+11.6}_{-1.1}$     \\
$\cdots$     &      &  8.6$^{+8.4}_{-2.5}$    & 7.2$^{+6.1}_{-3.7}$    & 2.4$^{+2.7}_{-1.7}$    & 10.1$^{+6.6}_{-0.0}$ & 6.3$^{+5.5}_{-3.5}$    & 15.5$^{+11.9}_{-10.1}$ & 2.8$^{+1.8}_{-0.0}$    & 8.5$^{+9.4}_{-0.0}$  & 33.5$^{+18.2}_{-9.2}$  & 5.0$^{+4.6}_{-0.0}$      \\
FQTau        & M2   & 17.9$^{+15.3}_{-11.8}$ & 1.0$^{+7.1}_{-0.3}$    & 0.4$^{+3.3}_{-0.2}$    & 5.2$^{+7.0}_{-2.7}$  & 49.6$^{+32.6}_{-28.2}$ & 9.1$^{+13.6}_{-4.1}$   & 3.0$^{+8.0}_{-2.4}$    & 4.8$^{+7.8}_{-3.0}$  & 9.0$^{+7.0}_{-3.6}$    & 0.0$^{+2.7}_{-0.0}$      \\
FSTau        & M1   & 32.5$^{+10.9}_{-9.6}$  & 1.6$^{+3.0}_{-0.6}$    & 0.3$^{+1.4}_{-0.0}$    & 1.3$^{+1.7}_{-0.6}$  & 42.6$^{+11.4}_{-14.7}$ & 0.0$^{+2.1}_{-0.0}$    & 0.0$^{+2.2}_{-0.0}$    & 0.0$^{+2.5}_{-0.0}$  & 17.0$^{+5.8}_{-6.1}$   & 4.6$^{+5.2}_{-3.2}$      \\
FTTau        & c    & 20.1$^{+5.5}_{-3.9}$   & 1.0$^{+1.4}_{-0.8}$    & 2.9$^{+1.8}_{-1.5}$    & 0.0$^{+1.0}_{-0.0}$  & 13.9$^{+8.6}_{-3.6}$   & 3.9$^{+3.4}_{-1.6}$    & 2.1$^{+2.1}_{-1.4}$    & 0.1$^{+1.6}_{-0.0}$  & 49.7$^{+15.9}_{-27.4}$ & 6.4$^{+10.3}_{-3.6}$     \\
$\cdots$     &      & 14.4$^{+17.8}_{-0.0}$  & 0.0$^{+10.9}_{-0.0}$   & 0.0$^{+4.4}_{-0.0}$    & 2.9$^{+13.7}_{-0.0}$ & 48.8$^{+22.9}_{-5.6}$  & 0.0$^{+23.7}_{-0.0}$   & 2.9$^{+11.9}_{-0.0}$   & 0.5$^{+35.3}_{-0.0}$ & 12.0$^{+30.6}_{-0.0}$  & 18.6$^{+15.1}_{-0.0}$    \\
FVTau        & K5   &  1.4$^{+6.5}_{-0.0}$    & 0.0$^{+4.8}_{-0.0}$    & 3.9$^{+3.0}_{-0.0}$    & 0.0$^{+0.8}_{-0.0}$  & 22.7$^{+19.4}_{-0.0}$  & 12.8$^{+8.7}_{-7.2}$   & 9.0$^{+11.8}_{-0.0}$   & 0.0$^{+3.0}_{-0.0}$  & 43.2$^{+12.9}_{-13.5}$ & 6.9$^{+16.1}_{-0.0}$     \\
FXTau        & M1   & 28.2$^{+8.6}_{-5.6}$   & 3.7$^{+5.0}_{-0.0}$    & 3.4$^{+1.8}_{-1.6}$    & 0.0$^{+0.8}_{-0.0}$  & 46.5$^{+10.5}_{-13.3}$ & 6.3$^{+4.2}_{-2.5}$    & 0.0$^{+2.1}_{-0.0}$    & 0.0$^{+1.6}_{-0.0}$  & 4.0$^{+3.6}_{-0.0}$    & 8.0$^{+2.6}_{-1.8}$      \\
$\cdots$     &      & 21.1$^{+14.8}_{-3.5}$  & 0.0$^{+8.8}_{-0.0}$    & 3.9$^{+8.7}_{-0.0}$    & 4.6$^{+16.3}_{-0.0}$ & 25.3$^{+16.6}_{-8.6}$  & 0.2$^{+2.7}_{-0.0}$    & 0.2$^{+4.2}_{-0.0}$    & 1.7$^{+10.7}_{-0.0}$ & 33.3$^{+16.6}_{-14.1}$ & 9.6$^{+21.9}_{-0.0}$     \\
FZTau        & M0   & 24.8$^{+7.5}_{-10.9}$  & 1.3$^{+2.2}_{-0.0}$    & 1.9$^{+1.0}_{-0.9}$    & 10.7$^{+2.6}_{-2.6}$ & 19.1$^{+10.8}_{-5.4}$  & 4.4$^{+2.3}_{-1.9}$    & 1.5$^{+2.2}_{-0.5}$    & 0.0$^{+2.8}_{-0.0}$  & 25.0$^{+6.4}_{-8.4}$   & 11.1$^{+4.0}_{-4.2}$     \\
GGTau        & M0   &  9.2$^{+3.4}_{-3.7}$    & 0.9$^{+2.0}_{-0.0}$    & 1.6$^{+1.4}_{-0.0}$    & 0.0$^{+0.9}_{-0.0}$  & 54.2$^{+15.5}_{-15.3}$ & 0.0$^{+2.1}_{-0.0}$    & 1.9$^{+2.8}_{-1.3}$    & 0.0$^{+1.1}_{-0.0}$  & 16.5$^{+4.7}_{-3.8}$   & 15.7$^{+4.9}_{-4.4}$     \\
GHTau        & M2   & 10.6$^{+3.1}_{-3.9}$   & 1.7$^{+1.7}_{-0.9}$    & 2.1$^{+1.1}_{-1.3}$    & 0.7$^{+1.4}_{-0.7}$  & 27.3$^{+15.0}_{-11.6}$ & 7.3$^{+3.2}_{-4.0}$    & 0.8$^{+2.7}_{-0.8}$    & 4.1$^{+3.1}_{-2.9}$  & 45.4$^{+26.6}_{-14.6}$ & 0.0$^{+4.3}_{-0.0}$      \\
$\cdots$     &      &  8.7$^{+4.5}_{-4.5}$    & 2.0$^{+4.1}_{-2.4}$    & 4.9$^{+5.4}_{-4.4}$    & 2.7$^{+2.7}_{-2.3}$   & 11.5$^{+8.2}_{-6.5}$   & 4.4$^{+4.1}_{-3.2}$    & 1.0$^{+1.8}_{-1.0}$    & 12.6$^{+10.3}_{-8.0}$ & 32.4$^{+19.8}_{-14.5}$ & 19.7$^{+18.4}_{-13.8}$ \\
GITau        & K6   & 35.6$^{+10.0}_{-10.5}$ & 0.4$^{+3.0}_{-0.0}$    & 1.4$^{+1.3}_{-0.9}$    & 0.0$^{+0.6}_{-0.0}$  & 39.3$^{+14.0}_{-7.7}$  & 5.7$^{+2.6}_{-2.7}$    & 1.6$^{+2.0}_{-1.0}$    & 1.0$^{+1.8}_{-0.0}$  & 6.4$^{+1.8}_{-1.8}$    & 8.4$^{+2.9}_{-2.1}$      \\
$\cdots$     &      & 16.0$^{+10.5}_{-7.3}$  & 8.4$^{+5.7}_{-3.8}$    & 11.1$^{+6.0}_{-5.2}$   & 1.9$^{+5.1}_{-1.9}$  & 20.5$^{+10.5}_{-7.6}$  & 0.0$^{+2.1}_{-0.0}$    & 0.0$^{+2.7}_{-0.0}$    & 0.1$^{+2.8}_{-0.0}$  & 33.9$^{+14.4}_{-12.7}$ & 8.1$^{+9.2}_{-4.9}$      \\
GKTau        & K7   & 64.5$^{+7.3}_{-10.7}$  & 0.0$^{+0.9}_{-0.0}$    & 3.1$^{+1.5}_{-1.3}$    & 0.0$^{+0.7}_{-0.0}$  & 18.0$^{+3.5}_{-6.6}$   & 3.7$^{+0.8}_{-1.4}$    & 0.0$^{+0.7}_{-0.0}$    & 1.7$^{+0.9}_{-0.7}$  & 7.3$^{+1.8}_{-1.8}$    & 1.8$^{+0.8}_{-0.8}$      \\
GMAur        & K3   & 37.3$^{+9.4}_{-11.2}$  & 0.0$^{+3.2}_{-0.0}$    & 3.1$^{+2.8}_{-0.0}$    & 0.0$^{+1.3}_{-0.0}$  & 16.0$^{+5.5}_{-4.1}$   & 12.2$^{+3.4}_{-3.5}$   & 14.9$^{+5.6}_{-4.2}$   & 0.0$^{+1.2}_{-0.0}$  & 1.2$^{+10.8}_{-0.0}$   & 15.4$^{+4.5}_{-6.7}$     \\
$\cdots$     &      & 15.7$^{+29.0}_{-0.0}$  & 0.7$^{+5.5}_{-0.0}$    & 0.4$^{+29.4}_{-0.0}$   & 17.3$^{+33.6}_{-0.0}$ & 9.6$^{+14.1}_{-0.0}$   & 3.8$^{+20.4}_{-0.0}$   & 0.0$^{+8.6}_{-0.0}$    & 17.4$^{+17.6}_{-0.0}$ & 33.6$^{+32.2}_{-0.0}$  & 1.3$^{+14.1}_{-0.0}$   \\
GOTau        & M0   & 12.0$^{+6.5}_{-7.6}$   & 0.0$^{+4.8}_{-0.0}$    & 3.4$^{+2.8}_{-2.3}$    & 0.1$^{+2.0}_{-0.0}$  & 63.2$^{+24.7}_{-37.8}$ & 3.3$^{+6.0}_{-4.1}$    & 0.0$^{+5.4}_{-1.2}$    & 5.6$^{+7.8}_{-4.8}$  & 12.4$^{+6.4}_{-5.3}$    & 0.0$^{+2.0}_{-0.2}$      \\
HKTau        & M1   &  1.0$^{+2.9}_{-0.3}$    & 0.4$^{+2.0}_{-0.0}$    & 1.8$^{+1.3}_{-0.0}$    & 0.1$^{+1.1}_{-0.0}$  & 37.5$^{+6.1}_{-21.5}$  & 7.7$^{+2.8}_{-4.7}$    & 8.2$^{+4.0}_{-2.8}$    & 0.0$^{+3.3}_{-0.0}$  & 18.6$^{+8.1}_{-6.0}$   & 24.8$^{+4.5}_{-11.5}$    \\
$\cdots$     &      & 19.5$^{+29.6}_{-3.4}$  & 5.3$^{+10.3}_{-0.0}$   & 1.3$^{+8.8}_{-0.0}$    & 11.1$^{+54.3}_{-2.5}$ & 13.5$^{+33.1}_{-3.4}$  & 6.7$^{+33.1}_{-0.0}$   & 2.9$^{+5.6}_{-0.0}$    & 7.4$^{+64.7}_{-0.0}$ & 21.5$^{+25.3}_{-6.3}$  & 10.6$^{+14.0}_{-6.4}$   \\
HNTau        & K5   & 13.9$^{+6.7}_{-3.7}$   & 0.0$^{+0.4}_{-0.0}$    & 0.8$^{+0.7}_{-0.0}$    & 0.0$^{+0.4}_{-0.0}$  & 45.5$^{+12.1}_{-7.7}$  & 0.0$^{+1.3}_{-0.0}$    & 0.0$^{+1.3}_{-0.0}$    & 0.0$^{+1.3}_{-0.0}$  & 22.9$^{+4.8}_{-4.6}$   & 16.8$^{+4.5}_{-3.2}$     \\
HOTau        & M0.5 & 10.8$^{+3.7}_{-1.6}$   & 0.7$^{+1.0}_{-0.4}$    & 2.0$^{+2.6}_{-0.6}$    & 0.0$^{+0.7}_{-0.0}$  & 34.6$^{+12.9}_{-8.6}$  & 6.2$^{+3.6}_{-3.9}$    & 2.2$^{+5.5}_{-0.0}$    & 0.4$^{+2.2}_{-0.0}$  & 43.0$^{+11.7}_{-17.1}$ & 0.0$^{+10.3}_{-0.0}$     \\
$\cdots$     &      & 38.3$^{+19.0}_{-16.0}$ & 0.0$^{+1.2}_{-0.0}$    & 1.6$^{+4.0}_{-0.0}$    & 0.0$^{+2.7}_{-0.0}$  & 37.6$^{+23.2}_{-11.3}$ & 0.4$^{+5.9}_{-0.0}$    & 0.3$^{+2.0}_{-0.0}$    & 4.2$^{+7.1}_{-3.4}$  & 8.3$^{+7.0}_{-3.6}$    & 9.3$^{+3.2}_{-5.8}$      \\
HPTau        & K3   & 57.4$^{+10.9}_{-10.6}$ & 0.0$^{+0.8}_{-0.0}$    & 1.0$^{+1.0}_{-0.0}$    & 0.0$^{+0.9}_{-0.0}$  & 22.1$^{+3.6}_{-4.9}$   & 1.1$^{+1.9}_{-0.0}$    & 0.0$^{+1.3}_{-0.0}$    & 0.9$^{+1.3}_{-0.0}$  & 14.3$^{+4.4}_{-2.3}$   & 3.4$^{+1.3}_{-1.3}$      \\
Haro6-37     & K7   & 11.4$^{+13.5}_{-7.7}$  & 0.0$^{+1.6}_{-0.0}$    & 3.4$^{+2.2}_{-1.8}$    & 3.0$^{+2.9}_{-1.8}$  & 45.0$^{+16.2}_{-27.5}$ & 10.4$^{+5.8}_{-5.4}$   & 2.6$^{+4.5}_{-0.0}$    & 7.0$^{+6.5}_{-4.0}$  & 17.2$^{+12.2}_{-7.0}$  & 0.0$^{+7.1}_{-0.0}$      \\
IPTau        & M0   & 28.0$^{+9.2}_{-7.4}$   & 0.8$^{+2.4}_{-0.3}$    & 2.5$^{+1.4}_{-1.7}$    & 0.0$^{+0.5}_{-0.0}$  & 33.0$^{+9.9}_{-7.5}$   & 4.5$^{+4.4}_{-2.4}$    & 3.9$^{+5.0}_{-2.5}$    & 0.0$^{+1.2}_{-0.0}$  & 18.7$^{+12.5}_{-8.3}$  & 8.5$^{+3.2}_{-4.1}$      \\
$\cdots$     &      & 17.0$^{+8.1}_{-6.0}$   & 0.2$^{+7.4}_{-0.0}$    & 1.7$^{+5.7}_{-0.1}$    & 8.2$^{+5.6}_{-2.9}$  & 19.3$^{+18.2}_{-8.9}$  & 4.2$^{+18.2}_{-0.0}$   & 1.8$^{+9.0}_{-0.0}$    & 13.7$^{+8.5}_{-5.7}$ & 32.3$^{+11.1}_{-6.9}$  & 1.7$^{+12.0}_{-1.4}$     \\
IQTau        & M0.5 & 17.3$^{+4.1}_{-4.0}$   & 0.0$^{+1.4}_{-0.0}$    & 2.0$^{+1.7}_{-1.1}$    & 0.0$^{+0.9}_{-0.0}$  & 30.7$^{+8.8}_{-8.7}$   & 8.4$^{+3.6}_{-2.4}$    & 2.7$^{+3.5}_{-1.4}$    & 1.6$^{+2.4}_{-0.9}$  & 28.7$^{+6.8}_{-11.3}$  & 8.7$^{+3.9}_{-3.8}$      \\
$\cdots$     &      & 24.3$^{+8.9}_{-6.3}$   & 0.0$^{+1.2}_{-0.0}$    & 0.7$^{+2.7}_{-0.4}$    & 1.6$^{+3.1}_{-1.0}$  & 43.9$^{+14.6}_{-12.9}$ & 0.0$^{+4.6}_{-0.0}$    & 0.6$^{+3.1}_{-0.0}$    & 0.0$^{+6.5}_{-0.0}$  & 14.7$^{+9.1}_{-3.3}$   & 14.1$^{+12.4}_{-7.1}$    \\
ISTau        & K7   &  0.0$^{+4.1}_{-0.0}$    & 0.0$^{+3.3}_{-0.0}$    & 6.8$^{+2.9}_{-1.8}$    & 11.6$^{+3.8}_{-3.3}$ & 47.8$^{+13.5}_{-16.1}$ & 19.0$^{+7.3}_{-5.2}$   & 5.1$^{+7.1}_{-4.2}$    & 0.3$^{+4.4}_{-0.0}$  & 9.4$^{+8.4}_{-3.0}$    & 0.0$^{+7.7}_{-0.0}$      \\
$\cdots$     &      &  3.0$^{+24.8}_{-0.0}$   & 0.0$^{+31.7}_{-0.0}$   & 10.4$^{+55.5}_{-0.0}$  & 6.8$^{+66.0}_{-0.0}$ & 26.5$^{+42.8}_{-13.3}$ & 0.0$^{+76.9}_{-0.0}$   & 14.0$^{+97.4}_{-0.0}$  & 4.7$^{+66.6}_{-0.0}$ & 26.9$^{+67.9}_{-6.9}$  & 7.6$^{+47.0}_{-0.0}$     \\
LkCa15       & K5   & 55.9$^{+11.0}_{-15.1}$ & 0.0$^{+2.0}_{-0.0}$    & 2.7$^{+3.8}_{-0.0}$    & 0.0$^{+0.6}_{-0.0}$  & 21.6$^{+11.5}_{-6.1}$  & 7.7$^{+2.7}_{-2.7}$    & 6.3$^{+3.5}_{-3.0}$    & 0.0$^{+0.6}_{-0.0}$  & 4.8$^{+2.7}_{-0.0}$    & 0.8$^{+3.1}_{-0.0}$      \\
$\cdots$     &      & 22.0$^{+11.7}_{-8.4}$  & 9.2$^{+6.1}_{-6.1}$    & 0.0$^{+2.0}_{-0.0}$    & 10.2$^{+7.4}_{-5.8}$ & 16.0$^{+7.7}_{-6.8}$   & 6.1$^{+5.6}_{-3.6}$    & 0.0$^{+1.5}_{-0.4}$    & 6.6$^{+6.3}_{-4.9}$  & 16.2$^{+8.1}_{-6.9}$   & 13.7$^{+9.5}_{-7.0}$     \\
RWAur        & K3   & 26.1$^{+9.0}_{-9.6}$   & 0.3$^{+1.0}_{-0.0}$    & 1.5$^{+2.1}_{-0.0}$    & 0.0$^{+0.5}_{-0.0}$  & 39.0$^{+9.1}_{-23.1}$  & 2.3$^{+3.9}_{-0.0}$    & 0.0$^{+1.8}_{-0.0}$    & 0.0$^{+1.5}_{-0.0}$  & 12.4$^{+3.3}_{-8.7}$   & 18.3$^{+6.7}_{-6.8}$     \\
RYTau        & G1   &  6.5$^{+3.8}_{-1.8}$    & 8.4$^{+1.7}_{-2.7}$    & 1.1$^{+0.8}_{-0.6}$    & 8.0$^{+1.7}_{-1.5}$  & 69.7$^{+13.3}_{-10.2}$ & 1.9$^{+3.3}_{-0.0}$    & 0.0$^{+1.6}_{-0.0}$    & 1.0$^{+2.1}_{-0.0}$  & 3.3$^{+1.8}_{-1.1}$    & 0.0$^{+0.2}_{-0.0}$      \\
SUAur        & G1   & 20.2$^{+5.0}_{-5.4}$   & 0.0$^{+0.6}_{-0.0}$    & 1.3$^{+1.0}_{-0.9}$    & 0.0$^{+0.6}_{-0.0}$  & 49.7$^{+8.5}_{-8.5}$   & 1.2$^{+1.8}_{-0.0}$    & 2.4$^{+3.3}_{-0.0}$    & 0.0$^{+0.6}_{-0.0}$  & 3.9$^{+1.9}_{-1.7}$    & 21.3$^{+3.1}_{-3.6}$     \\
UYAur        & K7   & 19.9$^{+6.4}_{-6.8}$   & 0.1$^{+2.6}_{-0.0}$    & 0.0$^{+1.3}_{-0.0}$    & 0.0$^{+1.9}_{-0.0}$  & 54.3$^{+11.1}_{-13.1}$ & 0.0$^{+1.9}_{-0.0}$    & 1.3$^{+3.6}_{-0.0}$    & 0.0$^{+1.3}_{-0.0}$  & 14.1$^{+3.9}_{-6.2}$   & 10.4$^{+7.2}_{-0.0}$     \\
V710Tau      & M1   & 27.7$^{+14.5}_{-19.2}$ & 0.0$^{+1.6}_{-0.0}$    & 2.8$^{+2.6}_{-1.8}$    & 0.9$^{+1.45}_{-0.6}$ & 39.1$^{+17.8}_{-28.1}$ & 8.5$^{+6.9}_{-5.0}$    & 6.0$^{+5.9}_{-3.5}$    & 8.0$^{+6.2}_{-4.6}$  & 3.7$^{+1.9}_{-1.8}$    & 3.3$^{+2.5}_{-1.3}$      \\
V773Tau      & K3   & 26.2$^{+24.6}_{-5.4}$   & 3.8$^{+5.9}_{-0.0}$    & 1.3$^{+2.3}_{-0.0}$    & 0.0$^{+1.8}_{-0.0}$  & 40.9$^{+12.7}_{-17.2}$ & 4.3$^{+9.4}_{-0.0}$    & 0.0$^{+1.8}_{-0.0}$    & 0.0$^{+8.0}_{-0.0}$  & 10.9$^{+5.7}_{-3.8}$   & 12.5$^{+5.6}_{-6.3}$     \\
V836Tau      & K7   &  1.2$^{+3.4}_{-0.0}$    & 0.9$^{+0.8}_{-0.6}$    & 4.5$^{+1.9}_{-0.8}$    & 0.0$^{+0.9}_{-0.0}$  & 32.4$^{+16.9}_{-14.5}$ & 9.1$^{+4.2}_{-3.3}$    & 2.8$^{+3.4}_{-1.9}$    & 3.9$^{+2.1}_{-2.8}$  & 45.1$^{+9.7}_{-35.6}$  & 0.0$^{+8.7}_{-0.0}$      \\
$\cdots$     &      & 30.7$^{+48.4}_{-8.9}$   & 0.0$^{+26.6}_{-0.0}$   & 1.7$^{+9.2}_{-0.0}$    & 10.4$^{+61.5}_{-0.0}$ & 32.6$^{+52.5}_{-19.6}$ & 3.9$^{+23.7}_{-0.0}$   & 1.7$^{+12.2}_{-0.0}$   & 0.5$^{+24.0}_{-0.0}$ & 17.0$^{+30.6}_{-3.8}$  & 1.5$^{+3.7}_{-1.5}$     \\
V955Tau      & K5   &  0.0$^{+1.3}_{-0.0}$    & 0.0$^{+0.9}_{-0.0}$    & 6.3$^{+2.5}_{-1.6}$    & 10.7$^{+3.1}_{-2.8}$ & 59.3$^{+15.0}_{-32.6}$ & 4.6$^{+5.8}_{-2.1}$    & 0.0$^{+3.0}_{-0.0}$    & 0.0$^{+2.9}_{-0.1}$  & 11.9$^{+4.7}_{-3.6}$   & 7.2$^{+6.7}_{-1.8}$      \\
VYTau        & M0   &  2.5$^{+13.5}_{-0.0}$   & 0.0$^{+12.4}_{-0.0}$   & 3.3$^{+16.1}_{-0.0}$   & 0.0$^{+8.4}_{-0.0}$  & 18.1$^{+4.7}_{-8.2}$   & 1.1$^{+7.2}_{-0.1}$    & 0.0$^{+14.7}_{-0.0}$   & 0.6$^{+7.6}_{-0.0}$  & 74.0$^{+14.1}_{-47.6}$ & 0.3$^{+18.0}_{-0.0}$     \\
ZZTauIRS     & M4.5 & 16.0$^{+6.3}_{-3.5}$   & 0.7$^{+3.0}_{-0.0}$    & 0.7$^{+2.2}_{-0.0}$    & 0.1$^{+1.2}_{-0.0}$  & 76.1$^{+12.9}_{-21.8}$ & 0.3$^{+4.3}_{-0.5}$    & 3.3$^{+4.8}_{-1.9}$    & 0.0$^{+2.9}_{-0.0}$  & 0.0$^{+1.6}_{-0.0}$    & 2.8$^{+7.3}_{-0.0}$      \\
$\cdots$     &      & 16.1$^{+8.6}_{-6.5}$   & 0.0$^{+2.0}_{-0.2}$    & 8.1$^{+6.9}_{-5.0}$    & 6.0$^{+4.7}_{-3.3}$  & 39.4$^{+18.7}_{-15.4}$ & 0.0$^{+3.7}_{-0.4}$    & 9.1$^{+7.5}_{-5.2}$    & 9.7$^{+9.3}_{-6.1}$  & 7.0$^{+5.2}_{-3.5}$    & 4.6$^{+4.3}_{-2.9}$      \\

\hline
\end{tabular}
\flushleft{$^a$ Amorphous olivine and pyroxene combined.} \\
$^b$ For each object, the first line corresponds to the warm component
abundances, while the second line corresponds to the cold component
abundances, when available.\\
\end{table}

\begin{table}
\centering
\tiny
\caption{Continuation} 
\begin{tabular}{lc|cccc|cccc|cc}
\hline
ID & SpT & Oli/Pyr$^a$ \% & Ens \% & For \% & Sil \% & Oli/Pyr \% & Ens \% & For \% & Sil \% & Oli/Pyr \% & Sil \% \\
   &     & (0.1 $\mu$m) & (0.1 $\mu$m) & (0.1 $\mu$m) & (0.1 $\mu$m) & (1.5 $\mu$m) & (1.5 $\mu$m) & (1.5 $\mu$m) & (1.5 $\mu$m) & (6.0 $\mu$m) & (6.0 $\mu$m) \\
\hline
\multicolumn{12}{c}{Upper Scorpius}  \\
PBB2002 J160357.9 & M2   & 8.2$^{+16.3}_{-2.1}$   & 0.0$^{+1.5}_{-0.0}$   & 2.1$^{+4.6}_{-0.2}$  & 0.0$^{+1.0}_{-0.0}$  & 42.9$^{+14.8}_{-24.5}$ & 0.0$^{+0.5}_{-4.5}$   & 0.0$^{+4.7}_{-0.0}$  & 0.0$^{+3.9}_{-0.0}$  & 46.8$^{+28.8}_{-32.4}$ & 0.0$^{+10.9}_{-0.0}$ \\
PBB2002 J160823.2 & K9   & 2.4$^{+6.2}_{-0.0}$    & 0.0$^{+5.3}_{-0.0}$   & 3.8$^{+6.2}_{-0.0}$  & 0.0$^{+1.1}_{-0.0}$  & 27.1$^{+2.5}_{-24.1}$  & 5.9$^{+3.9}_{-3.6}$   & 0.0$^{+7.7}_{-0.0}$  & 0.6$^{+10.4}_{-0.0}$ & 60.1$^{+38.8}_{-22.7}$ & 0.0$^{+5.1}_{-0.1}$  \\
PBB2002 J160900.7 & K9   & 17.2$^{+6.5}_{-4.5}$   & 0.5$^{+1.4}_{-0.0}$   & 3.0$^{+2.0}_{-1.6}$  & 0.0$^{+0.3}_{-0.0}$  & 42.3$^{+14.3}_{-17.3}$ & 4.1$^{+3.7}_{-2.4}$   & 0.0$^{+2.5}_{-0.0}$  & 0.0$^{+1.0}_{-0.0}$  & 24.9$^{+2.9}_{-16.8}$  & 8.0$^{+4.9}_{-4.0}$  \\
PBB2002 J160959.4 & M4   & 17.7$^{+31.6}_{-11.4}$ & 0.0$^{+2.7}_{-0.0}$   & 4.8$^{+3.8}_{-2.4}$  & 0.5$^{+2.5}_{-0.0}$  & 61.9$^{+16.5}_{-50.2}$ & 10.6$^{+12.0}_{-8.1}$ & 0.0$^{+7.6}_{-0.0}$  & 0.0$^{+4.5}_{-0.0}$  & 3.1$^{+5.9}_{-0.9}$    & 1.2$^{+11.7}_{-0.2}$ \\
PBB2002 J161115.3 & M1   & 20.3$^{+4.2}_{-12.7}$  & 0.0$^{+2.4}_{-0.0}$   & 1.9$^{+2.5}_{-1.4}$  & 0.0$^{+0.7}_{-0.0}$  & 58.7$^{+11.1}_{-26.8}$ & 3.1$^{+5.0}_{-2.0}$   & 0.0$^{+6.7}_{-0.0}$  & 1.5$^{+2.6}_{-0.4}$  & 9.2$^{+3.2}_{-4.6}$    & 5.3$^{+1.3}_{-4.9}$  \\
PBB2002 J161420.2 & M0   & 6.5$^{+1.4}_{-1.4}$    & 0.0$^{+0.5}_{-0.0}$   & 1.4$^{+0.9}_{-0.9}$  & 0.0$^{+0.3}_{-0.0}$  & 67.7$^{+7.9}_{-7.8}$   & 0.6$^{+0.9}_{-0.0}$   & 0.8$^{+1.2}_{-0.0}$  & 0.0$^{+0.4}_{-0.0}$  & 15.2$^{+2.8}_{-1.5}$   & 7.9$^{+1.3}_{-1.3}$  \\
PZ99 J160357.6    & K5   & 0.0$^{+3.1}_{-0.0}$    & 0.0$^{+2.8}_{-0.0}$   & 2.8$^{+5.1}_{-0.0}$  & 0.0$^{+3.2}_{-0.0}$  & 6.1$^{+9.2}_{-0.0}$    & 3.8$^{+2.9}_{-0.0}$   & 0.2$^{+2.2}_{-0.0}$  & 0.0$^{+2.6}_{-0.0}$  & 87.1$^{+19.9}_{-46.7}$ & 0.0$^{+5.1}_{-0.0}$  \\
PZ99 J161411.0    & K0   & 0.0$^{+2.0}_{-0.0}$    & 0.0 $^{+4.0}_{-0.0}$  & 1.1$^{+2.1}_{-0.0}$  & 1.4$^{+4.4}_{-0.0}$  & 8.8$^{+4.7}_{-0.0}$    & 3.9$^{+3.5}_{-0.0}$   & 0.0$^{+1.4}_{-0.0}$  & 0.0$^{+4.1}_{-0.0}$  & 84.8$^{+8.6}_{-55.2}$  & 0.0$^{+3.0}_{-0.0}$  \\
ScoPMS31          & M0.5 & 9.6$^{+7.5}_{-11.3}$   & 0.0$^{+2.5}_{-0.0}$   & 1.9$^{+3.3}_{-1.8}$  & 3.0$^{+4.6}_{-1.6}$  & 72.4$^{+39.8}_{-61.0}$ & 4.6$^{+10.7}_{-3.7}$  & 0.0$^{+9.0}_{-1.9}$  & 1.9$^{+9.2}_{-1.4}$  & 6.6$^{+4.2}_{-6.2}$     & 0.0$^{+1.5}_{-0.0}$  \\
\hline
\multicolumn{12}{c}{Eta Chamaeleontis}  \\
J0843   & M3.4 & 27.2$_{-11.0}^{+16.9}$ & 1.1$_{-0.0}^{+8.3}$ & 4.4$_{-0.0}^{+3.3}$  & 1.5$_{-0.0}^{+2.8}$  & 44.6$_{-21.9}^{+19.9}$ & 5.9$_{-0.0}^{+7.6}$   & 0.0$_{-0.0}^{+10.2}$ & 1.6$_{-0.0}^{+11.6}$ & 13.6$_{-0.0}^{+19.2}$ & 0.0$_{-0.0}^{+3.9}$  \\
RECX-5  & M3.8 & 36.1$_{-19.4}^{+23.4}$ & 0.0$_{-0.0}^{+2.9}$ & 12.3$_{-0.0}^{+8.7}$ & 0.3$_{-0.0}^{+14.8}$ & 12.3$_{-3.5}^{+15.6}$  & 23.9$_{-0.0}^{+13.9}$ & 0.0$_{-0.0}^{+8.0}$  & 8.0$_{-3.3}^{+6.9}$  & 7.1$_{-0.0}^{+18.9}$  & 0.0$_{-0.0}^{+10.0}$ \\
RECX-9  & M4.4 & 42.9$_{-34.1}^{+21.6}$ & 0.0$_{-0.0}^{+6.0}$ & 1.9$_{-1.0}^{+3.9}$  & 8.6$_{-6.7}^{+8.3}$  & 31.6$_{-24.2}^{+15.8}$ & 0.9$_{-0.0}^{+7.0}$   & 5.4$_{-3.2}^{+5.8}$  & 4.6$_{-2.1}^{+10.8}$ & 3.3$_{-1.1}^{+3.0}$   & 0.8$_{-0.0}^{+8.3}$  \\
RECX-11 & K6.5 & 36.9$_{-17.2}^{+18.1}$ & 1.4$_{-0.0}^{+3.4}$ & 3.4$_{-0.0}^{+8.4}$  & 2.6$_{-0.0}^{+4.2}$  & 26.8$_{-14.4}^{+15.7}$ & 8.0$_{-0.0}^{+10.5}$  & 0.0$_{-0.0}^{+2.7}$  & 3.4$_{-1.9}^{+3.7}$  & 17.6$_{-6.6}^{+8.1}$  & 0.0$_{-0.0}^{+1.9}$  \\

\hline
\end{tabular}
\flushleft{$^a$ Amorphous olivine and pyroxene combined.} \\
$^b$ For each object, the first line corresponds to the warm component
abundances, while the second line corresponds to the cold component
abundances, when available.\\
\end{table}

\end{document}